\documentclass[12pt,a4paper]{article}
\usepackage{amsmath}
\usepackage{amssymb}
\usepackage{graphicx}
\begin{document}

\title{Bubble nucleation and growth in slow cosmological phase transitions}
\author{\large  Ariel M\'{e}gevand\thanks{%
Member of CONICET, Argentina. E-mail address: megevand@mdp.edu.ar}~ and
Santiago Ram\'{\i}rez \\[0.5cm]
\normalsize \it IFIMAR (UNMdP-CONICET)\\ \normalsize \it Departamento de
F\'{\i}sica, Facultad de Ciencias Exactas y Naturales, \\ \normalsize \it
UNMdP, De\'{a}n Funes 3350, (7600) Mar del Plata, Argentina }
\date{}

\maketitle

\begin{abstract}
We study the dynamics of cosmological phase transitions in the case of small velocities of bubble walls, $v_w<0.1$. 
We discuss the conditions in which this scenario arises in a physical model, and
we compute the development of the phase transition.
We consider different kinds of approximations and refinements for 
relevant aspects of the dynamics, such as  
the dependence of the wall velocity on hydrodynamics, the distribution of the latent heat, and the variation of the nucleation rate.
Although in this case the common simplifications of a constant wall velocity and an exponential nucleation rate break down due to reheating, we show that a delta-function rate and a velocity which depends linearly on the temperature give a good description of the dynamics and allow to solve the evolution analytically.
We also consider a Gaussian nucleation rate, which gives a more precise result for the bubble size distribution.  
We  discuss the implications for the computation of cosmic remnants. 
\end{abstract}

\section{Introduction}

A first-order phase transition in the early universe causes disturbances
in the plasma which may result in the production of cosmic relics
such as topological defects \cite{k76}, gravitational waves \cite{tw90},
the baryon asymmetry of the universe \cite{krs85},
 or baryon inhomogeneities
\cite{w84,h95}. In a first-order phase transition, the system is
initially supercooled in a phase which is metastable below the critical temperature $T=T_c$. The transition to
the stable phase proceeds through the nucleation and expansion of
bubbles, and the properties of the generated cosmic relics
depend strongly on this dynamics. The relevant quantities are the bubble nucleation
rate per unit volume per unit time, $\Gamma$, and the velocity of
bubble walls, $v_{w}$. 

The nucleation rate vanishes for $T\geq T_{c}$ and, below $T_{c}$, it grows very rapidly as the temperature decreases. 
Bubble nucleation becomes appreciable when  $\Gamma$ becomes comparable
to the Hubble rate $H=\dot{a}/a$  \cite{tww92}, where $a$ is the
scale factor and a dot means a derivative with respect to time. 
Since $\Gamma$ is extremely sensitive
to the difference $T_{c}-T$, the most reliable  approach is to compute it numerically.  
However, in order to simplify the treatment of the dynamics, approximations are often used.
Since $\Gamma$ has the well-known form $\Gamma=Ae^{-S}$, where  $S$ (the instanton action) is a rapidly varying function of the temperature,
the most common approximation is to assume a constant factor $A$ and linearize $S(T(t))$ around a certain time $t_*$
This gives
an exponential rate $\Gamma(t)=\Gamma_*e^{\beta_*(t-t_*)}$.

The bubble wall velocity also vanishes at
$T=T_{c}$, since the pressure is the same in the two phases. At lower temperatures, the pressure is higher in the stable phase, and 
the walls move. Their velocity $v_w$ depends on the friction with the plasma, and is also affected by non-trivial hydrodynamics.
 Besides, the interactions between bubbles must be taken into account
in the global dynamics of the phase transition.
To simplify the treatment, a common approximation is to assume that $v_w$ remains constant during the phase transition. Since the pressure difference
goes roughly with the difference $T_{c}-T$,  this will be a reasonable approximation as long as the temperature has a relatively small variation, $\delta T\ll T_c-T$. In principle, 
the temperature decreases with time due to the adiabatic expansion of the universe, and this condition translates into a condition for the time intervals, $\delta t\ll t-t_c$.  
For an exponential nucleation rate, this requires $\beta_*^{-1}\ll H^{-1}$, which is satisfied in general.

In some calculations, more drastic simplifications are needed, such as considering a constant nucleation rate or assuming that all bubbles nucleate at the same time. 
These approximations are quite different from the exponential growth, and are in general regarded as rough approximations.
However, the exponential approximation is not always
valid either. For instance, for very strong phase transitions, the function $S(T)$ may
have a minimum at a certain temperature $T_{m}$, and the supercooling
may be such that this temperature is reached. In such a case $S(T)$
cannot be linearized, and a Gaussian approximation for $\Gamma(t)$ is more suitable \cite{nos} (besides, the phase transition becomes slow \cite{csw17}).
Nevertheless, in most cases the temperature is not close to the minimum of $S(T)$, and the latter may be assumed to be a monotonically increasing function of $T$. For decreasing $T$ we have, 
around a given time $t_*$, an exponentially growing rate.

However, these arguments do not take into account the fact that latent
heat is released as bubbles expand. This energy reheats the plasma, causing the
temperature to reapproach $T_c$. As a consequence, both $\Gamma$ and $v_w$ may decrease, slowing down the phase transition. 
The latent heat is released at the bubble walls, and
the reheating is in general inhomogeneous. This fact makes the general
treatment of the phase transition difficult, except in some special
cases. One of them is the case of detonation bubbles \cite{s82},
in which the velocity of the walls is so high that the fluid in
front of them remains unperturbed. In this case, the reheating occurs
only inside the bubbles, so in the old phase the nucleation rate grows
according to the adiabatic cooling. Thus, for detonations, an exponential nucleation rate is generally a good approximation, and the phase transition is  quick enough to assume a constant wall velocity.

In contrast, for smaller velocities, the wall propagates as a deflagration front \cite{gkkm84} and is preceded by a shock wave which carries away the
released energy. In this case, the temperature outside the bubbles
may be very inhomogeneous, and the nucleation rate may vary significantly
from region to region. Nevertheless, 
the shock waves propagate supersonically and 
the latent heat quickly distributes throughout space. 
For small enough wall velocities, 
we may assume a homogeneous temperature 
as bubbles nucleate and expand \cite{h95}. Once the rate at which latent heat is released exceeds the energy decrease due to the adiabatic expansion, the temperature will stop decreasing and start to grow.

This dynamics has been discussed, to different extents, in Refs.~\cite{h95,m04,ma05,kk86,m01,ms08}. 
The treatment in these works is either numerical or involves rough approximations. This is because,
even assuming a homogeneous temperature, 
its variation is related to that of the wall velocity and the nucleation rate
through non-trivial integro-differential equations. 
In contrast, in the detonation case, 
the  assumptions of a constant wall velocity and an exponential nucleation
rate 
simplify considerably the computations. Depending on other approximations, it is even possible to obtain the complete development of the phase transition analytically. The results are functions of a few free parameters, such as $v_w$ and $\beta_*$, which can be computed numerically for specific models, and one obtains a very precise description of the phase transition (see \cite{nos} for a recent discussion). 

Notice that, in the slow deflagration case, the temperature will generally have a minimum $T_m$ at a time $t_{m}$ separating the supercooling and reheating stages. 
Around this time, the exponent $S(T(t))$ is quadratic in $t-t_m$, and $\Gamma(t)$ should be well approximated by a Gaussian function. Moreover, it
was pointed out in Refs.~\cite{m04,ma05} that the time during which
$\Gamma$ is effectively active can be quite smaller than the total
duration of the phase transition. In such a case, even a delta function
may be a suitable approximation for $\Gamma(t)$. On the other hand, 
slow deflagrations will occur in general
in phase transitions with little supercooling. In such a case, it is
a good approximation to consider $v_{w}$ to first order in $T_{c}-T$.
In this paper we shall investigate this scenario. 
We shall discuss the validity of 
these approximations for the nucleation rate and the wall velocity, and we will show that they
simplify considerably the numerical treatment, 
even allowing to obtain analytic solutions.
These approximations depend on a few parameters, 
which, for a given model, can be estimated numerically with simple computations. 

The plan is the following. 
In the next section we study the general dynamics of a phase transition mediated by slow deflagrations, and we discuss the possible approximations for the distribution of the latent heat and the dependence of the wall velocity on the temperature.
In Sec.~\ref{approxrate} we consider several approximations for
the nucleation rate.
In particular, we show that assuming  an exponential rate until the time $t_m$ gives a good estimation of the temperature $T_m$. 
However, 
assuming a Gaussian nucleation rate gives a better approximation for the final bubble size distribution.
On the other hand,
assuming a simultaneous nucleation at $t=t_m$ is a good approximation for computing the 
evolution of $T$ and $v_w$.
We study this evolution for a delta-function rate in Sec.~\ref{anf}. 
In Sec.~\ref{cosmo} we discuss on cosmological applications of our treatment, and in Sec.~\ref{conclu}
we summarize our conclusions. 

\section{Development of a slow phase transition \label{dyn}}

\subsection{Effective potential and model parameters} \label{effpot}

To describe a first-order phase transition, we shall consider a  model
consisting of a scalar field $\phi$ with  a spontaneous symmetry-breaking tree-level potential of the form $V(\phi)=-(m^{2}/2)\phi^{2}+(\lambda_{0}/4)\phi^{4}$, and particles which
acquire their masses $m_i$ from the vacuum expectation value the field $\phi$. For $m_i/T\lesssim1$, the one-loop finite-temperature effective potential $V(\phi,T)$ has an
expansion in powers of $\phi/T$ (see, e.g., \cite{ah92}), 
\begin{equation}
V(\phi,T)=D(T^{2}-T_{0}^{2})\phi^{2}-ET\phi^{3}+\frac{\lambda}{4}\phi^{4}. \label{potpots}
\end{equation}
The coefficients in Eq.~(\ref{potpots}) depend on the parameters $m$ and $\lambda_{0}$ as well as on the particle content. 
These coefficients are constant, except for $\lambda$, which has a  logarithmic dependence on $T$. Since 
the temperature will not depart significantly from the critical temperature, we shall neglect this dependence. 
The complete free energy density is given by
\begin{equation}
\mathcal{F}(\phi,T)=\rho_{V}-\frac{\pi^{2}}{90}g_{\ast}T^{4}+V(\phi,T),\label{freeen}
\end{equation}
where the constant $\rho_{V}$ is the false-vacuum energy density and
the second term is the radiation component. 
This term  is proportional to the total number of degrees of freedom of the species in the plasma, $g_{\ast}$, while only a part $\Delta g$ of these degrees of freedom have strong enough couplings to $\phi$ and contribute to the term $V(\phi,T)$.

At high enough temperatures, $V(\phi,T)$ has an absolute minimum at $\phi=0\equiv \phi_{+}$, while at low enough temperatures the absolute minimum is at $\phi=\phi_-$, with
\begin{equation}
\frac{\phi_{-}(T)}{T}=\frac{3E}{2\lambda}\left[1+\sqrt{1-\frac{8\lambda D}{9E^{2}}\left(1-\frac{T_{0}^{2}}{T^{2}}\right)}\right].
\end{equation}
There is a temperature range in  which
these two minima coexist. The lower temperature in this range is  $T=T_0$, below which $\phi=0$ becomes a maximum. The critical temperature, defined by the equation $\mathcal{F}(\phi_+,T_c)=\mathcal{F}(\phi_-,T_c)$, is given by
\begin{equation}
\frac{T_{c}^{2}-T_{0}^{2}}{T_{c}^{2}}=\frac{E^{2}}{\lambda D}.\label{tct0}
\end{equation}
In this model, the dimensional parameter
$T_{0}$ determines the temperature scale of the phase transition. This parameter can be related to the 
zero-temperature minimum $v$, which is usually considered as the energy scale of the theory,
${2D}T_{0}^{2}={\lambda} v^{2}$. 
The two phases are characterized by the functions 
$\mathcal{F}_{\pm}(T)=\mathcal{F}(\phi_{\pm},T)$, from which we can derive thermodynamic quantities such as
the pressure $p_{\pm}=-\mathcal{F}_{\pm}$, the entropy
density $s_{\pm}=-d\mathcal{F_{\pm}}/dT$, and the energy
density $\rho_{\pm}=\mathcal{F}_{\pm}-Td\mathcal{F}_{\pm}/dT$.
Thus, in the high-temperature phase, we have $\rho_{+}=\rho_{V}+\rho_{R}$, where $\rho_{R}$ is the radiation
energy density, 
\begin{equation}
\rho_{R}=\frac{\pi^{2}}{30}g_{\ast}T^{4},\label{rhoR}
\end{equation}
and the energy difference between the two phases is given by $\rho_{+}-\rho_{-}=T\partial V/\partial T-V$. Its value at the critical temperature is the latent heat $L$.  In  this model  we have 
\begin{equation}
L/T_{c}^{4}=2D\left(\phi_{c}/T_{c}\right)^{2}-E\left(\phi_{c}/T_{c}\right)^{3}, 
\label{lat}
\end{equation}
where $\phi_c\equiv\phi_{-}(T_{c})$ is the jump of $\phi$ at the critical temperature, which is given by 
\begin{equation}
\phi_c/T_c=2E/\lambda. \label{fic}
\end{equation} 

For $E=0$ we have $T_{c}=T_{0}$ and $\phi_c=0$, which means that the phase transition is second order. 
The strength of the phase transition is usually measured by the parameter $\phi_c/T_c$.
For $\phi_{c}/T_{c}\gtrsim1$
the phase transition is said to be strongly first order, while for
$\phi_{c}/T_{c}\ll 1$ it is said to be weakly first order. 
Using dimensionless quantities such as $T/T_c$ and $\phi(T)/T$, we see that
the dynamics will depend mainly on 
the dimensionless constants $D,E,\lambda$ which determine the shape of the effective potential.  
In general, for $\phi_c/T_c\sim 1$, most of the relevant quantities, such as $L/T_{c}^{4}$, will be of order 1. However,
the dynamics of reheating will depend on the ratio $L/\rho_{Rc}$, where $\rho_{Rc}=\rho_R(T_c)$  (a value $L\ll \rho_{Rc}$ will not cause a significant temperature change in the plasma).
Since only a few of the degrees of freedom contribute to the free energy difference $V(\phi,T)$,  we expect in general $L\sim T_c^4$, while we have $\rho_{Rc}\sim g_*T_c^4$. Hence, we will have, typically, $L/\rho_{Rc}\sim1/g_*$. 
On the other hand,  the dynamics will have a dependency on the temperature scale through 
the Friedmann equation
\begin{equation}
H=\frac{\dot{a}}{a}=\sqrt{\frac{8\pi}{3}\frac{\rho}{M_{P}^{2}}},\label{hubble}
\end{equation}
where $M_{P}=1.22\times10^{19}\mathrm{GeV}$  is the Plank mass. 
The false-vacuum energy density  in this model is given by $\rho_{V}\simeq\lambda v^{4}/4$, and for $T\simeq T_c\sim v$ we have $\rho_V/\rho_R\sim \lambda/g_*$. For many physical models there is a large number of degrees of freedom in the plasma, and we have $\rho_{V}\ll\rho_{R}$ and $L\ll\rho_R$, so 
$\rho_-\simeq\rho_+\simeq\rho_R$.

For our general treatment we may regard the constants $v$, $g_*$, $D$, $E$, and $\lambda$ as free parameters. 
For specific computations we shall consider electroweak-scale values for $v$ and $g_*$, namely,
$v=250\mathrm{GeV}$, $g_{\ast}=100$, and we shall set 
the value of $\lambda$ by demanding a natural value 
for the scalar mass $m_{\phi}=\sqrt{2\lambda_0 v^{2}}$. Assuming $\lambda_0\simeq\lambda$, 
we choose $\lambda=0.125$, corresponding to $m_\phi\simeq v/2$.  
On the other hand, the parameter $E$ is generally smaller, since it  depends cubically on the couplings of $\phi$ with gauge fields.  Hence, according to Eq.~(\ref{fic}),
this model does not naturally give very strong phase transitions\footnote{In this model, we have a fluctuation-induced cubic term $-ET\phi^3$, which appears at finite temperature. If we considered a tree-level cubic term $-A\phi^3$,  we might have $\phi_c/T_c\gg 1$ (see, e.g., \cite{nos}). Since we are not interested in such strong phase transitions, the model (\ref{potpots}) has all the qualitative features we need.}.
Nevertheless, for $E\geq\lambda/2$ we have 
$\phi_{c}/T_{c}\geq1$. We shall  consider  the value  $E=0.075$, corresponding to
$\phi_{c}/T_{c}=1.2$. The coefficient $D$, in contrast, is quadratic in the couplings and involves a sum over all particle species, and we may have $D\sim 1$.
Typically, we have $E<\lambda<D$, and  Eq.~(\ref{tct0}) gives $T_{c}-T_{0}\ll T_{0}$. 
Notice that, according to Eq.~(\ref{lat}),
for $\phi_c/T_c\simeq 1$ and  $E\ll D$ we have $L/T_c^4\simeq 2D\sim 1$, as expected.
For these natural values, we expect a reheating $\Delta T/T\sim L/\rho_R\sim10^{-2}$. This competes with the amount of supercooling, since the phase transition will take place in the range $T_{0}<T<T_{c}$, and we have
$(T_{c}-T_{0})/T_{c}\simeq E^{2}/(2\lambda D)\sim10^{-2}$. 
We shall consider  a couple of values of the ratio $L/\rho_{Rc}$; namely,  $L/\rho_{Rc}=0.025$ (corresponding to $L\simeq0.82T_{c}^{4}$ and $D\simeq0.33$), and 
$L/\rho_{Rc}=0.05$ (corresponding to $L\simeq1.64T_{c}^{4}$ and $D\simeq0.62$).

\subsection{Initial nucleation and growth}

The temperature variation is governed by the adiabatic-expansion equation $ds/dt=-3Hs$. In our model, for the high-temperature phase we have $s_+\propto T^3$. Hence, before the phase transition the temperature decreases with a rate  
\begin{equation}
\frac1{T}\frac{dT}{dt}=-H. \label{tTsuperc}
\end{equation}
Between $T=T_{c}$ and $T=T_{0}$ the effective potential has a barrier between the minima, and  the phase transition may proceed by bubble nucleation. For $T\leq T_0$ the barrier separating the minima disappears, and the phase transition will proceed by spinodal decomposition. In the bubble-nucleation range, there is a probability of nucleating a bubble per unit volume per unit time given by 
\cite{affleck,linde}
\begin{equation}
\Gamma(T)=A(T)\,e^{-S(T)}, \label{gamma}
\end{equation}
where $S(T)=S_{3}(T)/T$, $A(T)=T^{4}\left[S_{3}(T)/(2\pi T)\right]^{3/2}$, and $S_{3}$ is the spherically symmetric extremum  of the three-dimensional instanton action \cite{linde}.
This extremum also gives the configuration of the nucleated bubble. 
We will solve 
numerically the equation for the instanton configuration by the undershoot-overshoot method (see \cite{nos} for details).

It is well known that the action $S_{3}$ diverges at $T=T_{c}$ and vanishes at $T=T_{0}$. 
Hence, the nucleation rate vanishes at the critical temperature and
reaches values $\Gamma\sim T^{4}$ as $T$ approaches the value $T_{0}$. The latter is, relatively, a huge rate, since the phase transition will occur, roughly, for $\Gamma\sim H^4$, and we have $T^4/H^4\sim M_P^4/T^4$ (which is generally large unless the scale $v$ is very close to the Planck scale).
Hence,
the phase transition will generally complete before reaching the spinodal decomposition temperature. 
We shall take the time $t_H$ defined by the equality $\Gamma=H^{4}$ as a reference time.  
The corresponding temperature $T_H$ is given by the
equation $S-3/2\log(S/2\pi)=4\log(T/H)$.

The time $t_{N}$ at which bubble nucleation effectively begins 
is usually defined by the condition that there is one bubble
in a Hubble volume, and generally we have $t_{N}>t_{H}$ (see \cite{nos}
for a recent discussion). During this supercooling stage, the number density of bubbles is given by  
\begin{equation}
n(t)=\int_{t_{c}}^{t}dt^{\prime}\Gamma(t^{\prime}), \label{tN}
\end{equation}
with the time-temperature relation given by Eqs.~(\ref{tTsuperc}) and (\ref{hubble}), 
with $\rho=\rho_{+}$.
This expression does not take into account the fraction of volume occupied by
bubbles, which at this time is negligible, and we do not include the dilution of the number density
as $a^{-3}$ either, since we have $\Delta a/a\sim \Delta T/T\ll1$.
Thus, the time $t_N$ is determined from the condition $nH^{-3}=1$.

A nucleated bubble grows due to the pressure difference between the
two phases. 
The motion of a bubble wall causes bulk fluid motions as well as temperature gradients in the plasma, which affect the propagation of the phase transition front.
It is well known that the treatment of hydrodynamics is quite simple for
the bag equation of state (EOS) \cite{s82}.
In our case, the free energy (\ref{freeen}) has exactly the bag form in the phase $+$, 
\begin{equation}
\mathcal{F}_{+}(T)=\varepsilon_{+}-a_{+}T^{4}/3,
\end{equation}
 where $\varepsilon_{+}=\rho_{V}$ and $a_{+}=\pi^{2}g_{\ast}/30$.
On the other hand, since we have little supercooling, for the phase $-$  we may use the approximation
\begin{equation}
\mathcal{F}_-(T)\simeq\mathcal{F}_-(T_c)+\left.\frac{d\mathcal{F}_-}{dT^4}\right|_{T_c}(T^4-T_c^4),
\label{fmelin}
\end{equation}
which also has the bag form. Taking into account the relations $\mathcal{F}_-(T_c)=\mathcal{F}_+(T_c)$ and $L=4T_c^4(d\mathcal{F}_-/dT^4-d\mathcal{F}_+/dT^4)|_{T_c}$, we may write 
\begin{equation}
\mathcal{F}_{-}(T)=\varepsilon_{-}-a_{-}T^{4}/3,
\end{equation}
with 
$\varepsilon_-=\varepsilon_+-L/4$ and $a_-=a_+-3L/(4T_c^4)$.
With this approximation, several results will depend only on
the variable 
\begin{equation}
\alpha(T)=\frac{L}{4a_+T^4}.
\end{equation}
This quantity is larger for stronger phase transitions and smaller for weaker ones\footnote{Indeed, we may write $\alpha=\frac{1}{4}\frac{\Delta\rho(T_c)}{\rho_R(T_c)} \frac{T_c^4}{T^4}$. Hence, it is proportional to the relative energy discontinuity and inversely proportional to the amount of supercooling.}.

The force which drives the motion of a bubble wall is not simply given by the pressure difference $p_{-}(T)-p_{+}(T)$ since, in the first place, the temperature varies across the wall.
Using a linear approximation for the temperature variation inside the wall, the driving force can be written as \cite{lm16} 
\begin{equation}
F_{\mathrm{dr}}=\frac{L}{4}(1-T_{+}^{2}T_{-}^{2}/T_{c}^{4}),\label{fdr}
\end{equation}
where $T_+$ and $T_-$ are the values of $T$ in front and behind the wall, respectively. 
The relation between $T_{+}$ and $T_{-}$ is obtained by energy-momentum conservation \cite{landau}. 
This relation involves also the values of the fluid velocity on each side of the wall. 
In the wall reference frame, we denote the magnitude of the incoming fluid
velocity by $v_{+}$ and that of the outgoing fluid velocity by $v_{-}$. 
For non-relativistic velocities we have
\begin{equation}
w_{-}v_{-}=w_{+}v_{+}, \qquad p_{-}=p_{+}, \label{landaunr}
\end{equation}
and  the bag EOS gives
\begin{equation}
\frac{v_{+}}{v_{-}}=
\frac{a_-T_{-}^4}{a_+T_{+}^{4}}=
1-3\alpha(T_+). \label{discnr}
\end{equation}
Notice, also, that we have the relation $a_-/a_+=1-3\alpha(T_c)$.

There is also a friction force, which arises as a consequence of 
the departures of the particles distributions from their equilibrium values inside the wall
(see, e.g., \cite{mp95}). A phenomenological approach to this force is often used, and  will be sufficient for our purposes. In the wall reference frame, the friction is modeled by a function of the average fluid velocity $\bar v$, and the latter is approximated by 
$\bar{v}=(v_{-}+v_{+})/2$ \cite{ikkl94}. 
In the non-relativistic case we have 
\begin{equation}\label{ffr}
F_{\mathrm{fr}}=-\eta\bar{v},
\end{equation}
where the friction coefficient $\eta$ is a free parameter which may be inferred by matching to the full microphysics computation.

The wall velocity is  obtained from the steady state condition $F_{\mathrm{dr}}+F_{\mathrm{fr}}=0$.
To solve this equation, we need  additional relations between  the fluid velocities $v_\pm$ and $v_w$. These relations are obtained from the fluid equations and the boundary conditions.
For an isolated bubble, the fluid is at rest far in front of the wall as well as far behind it (at the bubble center). For the small wall velocities we are interested in, which are subsonic with respect to the 
bubble center,
it turns out that we have a vanishing fluid velocity inside the bubble, i.e., $v_{-}=v_{w}$. Hence,
Eqs.~(\ref{fdr}-\ref{ffr}) give
\begin{equation}
v_{w}=\frac{L}{4\eta}\frac{T_{c}^{4}-T_+^{4}\sqrt{(a_{+}/a_{-})(1-3\alpha_+)}}{T_{c}^{4}(1-3\alpha_+/2)}, \label{vwnr}
\end{equation}
where $\alpha_+=\alpha(T_+)$. 
Inside the bubble we also have a constant temperature. 

Since the wall is subsonic with respect to the fluid in front of it, this fluid is affected by the wall motion. The fluid equations relate the variables next to the wall, $T_+,v_+$, to the boundary conditions far in front. It turns out that
the fluid profile ends in a discontinuity, a shock front which 
propagates supersonically in the  phase $+$. The temperature  falls from  
the value $T_{+}$ in front of the wall to
a value $T_{\mathrm{sh}}$ at the shock front. Beyond this discontinuity we have an unperturbed fluid at temperature $T$.  The matching conditions here are similar to those for the phase transition front. 
For small fluid velocities, the profile of the shock wave can be solved analytically. In this case, the temperature variation  is small, $(T_+-T_{\mathrm{sh}})/T_{\mathrm{sh}}\sim v_w^2$. Besides,
the velocity of the shock front is  $v_{\mathrm{sh}}\simeq c_{s}$, 
and the shock discontinuity becomes very weak, 
$T_{\mathrm{sh}}\simeq T$ 
(the error of this approximations is of order $e^{-1/v_{w}^{3}}$ \cite{k85}). Hence, we have $T_+=T+\mathcal{O}(v_w^2)$.

For $T_+\simeq T\simeq T_c$ and $L/\rho_{Rc}\ll 1$, we have
\begin{equation}
v_{w}\simeq\frac{L}{\eta}\frac{T_{c}-T}{T_{c}}. \label{vwsimple}
\end{equation}
We shall use the complete expression (\ref{vwnr}) for the numerical computations, but we will see that Eq.~(\ref{vwsimple}) is a good approximation in our case. The friction coefficient $\eta$ is not directly related to the parameters of the effective potential, and  we shall regard it as an independent free parameter.
According to Eq.~(\ref{vwsimple}),  for dimensionally natural values $\eta\sim T_{c}^{4}\sim L$ we will have $v_{w}\sim(T_c-T_0)/T_c\sim10^{-2}$ for our model parameters.
For specific computations we will use the value $\eta/T_{c}^{4}=0.5$. 

For $\eta/T_c^4\ll 1$, the non-relativistic approximations in Eqs.~(\ref{landaunr}-\ref{ffr}) will no longer be valid. In the general case, Eqs.~(\ref{discnr}) become 
\begin{eqnarray}
\frac{a_{-}T_-^4}{a_{+}T_+^4}&=&\frac{v_{+}\gamma _{+}^{2}}{v_{-}\gamma _{-}^{2}},
\label{discrel0} 
\\
v_{-}&=&\left( \frac{v_{+}\left( 1+\alpha_+ \right) }{2}+\frac{\frac{1}{3}
-\alpha_+}{2v_{+}}\right) \pm \sqrt{\left( \frac{v_{+}\left( 1+\alpha_+ \right)
}{2}+\frac{\frac{1}{3}-\alpha_+ }{2v_{+}}\right) ^{2}-\frac{1}{3}},
\label{discrel}
\end{eqnarray}
where $\gamma_{\pm}=1/\sqrt{1-v_\pm^2}$.
Several extrapolations of the friction force (\ref{ffr}) to the relativistic case have been proposed.   Some of them  \cite{ekns10,m13} take into account the fact that the  friction  may saturate in the ultra-relativistic limit and the wall may run away \cite{bm09}. However,
for a potential of the form (\ref{potpots})  the bubble wall cannot run away, and a reasonable extrapolation for the friction force is given by  
\begin{equation} 
F_{\mathrm{fr}}= -\eta\overline{\gamma v}, \label{ffrrel}
\end{equation}
where $\overline{\gamma v}=(v_+\gamma_++v_-\gamma_-)/2$ (for a recent discussion, see \cite{nos}).

We have different kinds of hydrodynamic solutions, corresponding to different branches of the $v_+$-$v_-$ relation. 
If the velocity of
the incoming flow  is lower than the speed of sound in the plasma, $c_s=1/\sqrt{3}$, we  have a  deflagration, while 
if the incoming flow is supersonic, we have a detonation (see e.g.\ \cite{lm11} for details). 
Although we are interested in the case of very slow deflagrations, for comparison we shall also consider the detonation case. 
A detonation wall moves supersonically with respect to the fluid in front of it. As a consequence,  this fluid is not affected by the wall, and we have $v_+=v_w$ and $T_{+}=T$, where
$T$ is the value of the temperature in the absence of any bubbles. 
Using these conditions in Eqs.~(\ref{fdr}), (\ref{discrel0}-\ref{discrel}) (with  the $+$ sign), and (\ref{ffrrel}), we may solve for $v_{w}$ as a function of $\eta$ and $\alpha$ (for more details, see, e.g., \cite{ms09}).
If $\alpha$ is too small
or the friction is too large, it turns out that there is no detonation solution. In
this case we have to consider a deflagration, which is compatible
with smaller velocities. 
To obtain a detonation, we shall  consider  the value $\eta/T_{c}^{4}=10^{-3}$.

\subsection{Global dynamics} \label{globdyn}

For isolated bubbles, the temperature $T$ is  determined by the adiabatic expansion,
\begin{equation}
T=T_{c}\,a_c/a , \label{adexp}
\end{equation} 
where $a_c\equiv a(T_c)$. We may assume that, as temperature varies, the walls instantly take the  terminal velocity $v_w(T)$ (see, e.g., \cite{h95}). 
Once bubbles begin to meet each other, the growth of a cluster of bubbles will depend on the motion of the uncollided walls.
In the case of detonations, the conditions in front of a wall are always the same as for an isolated bubble (namely, $v_+=v_w$, $T_+=T$), and we expect the results for the wall velocity to remain essentially unchanged. Also, the nucleation dynamics in the phase $+$ is not affected by the presence of previously nucleated bubbles.
Therefore, $v_w$ and $\Gamma$ are functions of the outside temperature $T$, which is given by Eq.~(\ref{adexp}).
In contrast, for deflagrations, the fluid in which a bubble expands is affected by shock fronts coming from other bubbles. 
The main effect is an increase of the temperature, which will decrease the wall velocity. Besides, the reheating will diminish the nucleation of new bubbles.

Initially, the isolated deflagration bubbles are contained inside larger spheres whose surfaces are the shock fronts. Each of these ``shock bubbles'' has a radius $R_{\mathrm{sh}}\simeq (c_{s}/v_w)R$, where  
$R$ is the radius of the bubble. To characterize the moment after which bubbles cannot be regarded as isolated, let us consider the time $t_0$ at which, in average, neighboring shock bubbles have just met each other. We may estimate this time by the condition that the average shock radius matches the average bubble separation $d$ (which depends on the bubble number density).
At this time,  the fraction of volume which is in the new phase is roughly $f_{-}(t_0)\sim (v_{w}/c_{s})^{3}$.
For slow walls, we will have $f_-\ll 1$, which means that most of the phase transition will occur with interacting bubbles. 
Nevertheless, in this case we have extremely weak shocks which softly change the boundary conditions for the wall motion. Hence, we may assume that
Eq.~(\ref{vwnr}) still applies locally,  with an inhomogeneous temperature $T_+$.

Once the shock
waves of a bubble have reached several other bubbles and bounced several times between two neighboring bubbles\footnote{See \cite{kk86} for an analytical description of this process in the (1+1)-dimensional case.}, we may assume that the released energy is homogeneously distributed. 
It will take the shocks a few times  $t_0$ to complete this homogenization.
For $v_{w}\ll c_{s}$ this will happen when the fraction of volume in the new phase is still very small. 
Hence, during this homogenization process the reheating will still be insignificant, and 
we may just assume a homogeneous distribution of the latent heat from the beginning
(we have checked these features with our numerical computations).
This approximately instant spreading of the released energy will continue during the whole phase transition, due to the velocity difference between walls and shocks.

Notice that the temperature cannot be homogeneous everywhere, since we have discontinuities at the phase transition fronts, given by Eqs.~(\ref{landaunr}-\ref{discnr}).  For the bag EOS,
the matching condition $p_{-}=p_{+}$ in Eq.~(\ref{landaunr}) gives  $a_{-}T_{-}^{4}/3-\varepsilon_{-}=a_{+}T_{+}^{4}/3-\varepsilon_{+}$.  Hence, we have $d(a_{-}T_{-}^{4})/dt=d(a_{+}T_{+}^{4})/dt$; i.e., the energy density has the same variation on both sides of the walls. 
This is consistent with the assumption of a homogeneous distribution of the released energy, and we may assume homogeneous temperatures $T_\pm$ in each phase.
To study the temperature variation, the most straightforward way is to consider the conservation of entropy, 
\begin{equation}
s_{-}(T_{-})f_{-}+s_{+}(T_{+})f_+=s_c a_c^3/a^{3},\label{varsmedia}
\end{equation}  
where $s_c=s_{+}(T_{c})$, and $f_\pm$ is the fraction of volume occupied by each phase. Since the system is out of equilibrium, the entropy is not exactly conserved. In the appendix we estimate the error of this approximation, which for the present case is negligible.
Like in the detonation case, we shall denote the homogeneous temperature  outside the bubbles by $T_+\equiv T$.
Since $f_+=1-f_-$, we have 
\begin{equation}
\frac{T^{3}}{T_{c}^{3}}=\frac{s_+(T)-s_-(T_-)}{s_{c}}f_-+\frac{a_c^3}{a^{3}}, \label{rec}
\end{equation}
and $T_-(T)$ is given by Eq.~(\ref{discnr}). 
The wall velocity is still given by Eq.~(\ref{vwnr}), with $T_+=T$ and $\alpha_+=\alpha$.

Thus, for very slow deflagrations the wall velocity and the nucleation rate depend on a single variable $T$, like in the detonation case. 
The main qualitative difference in the dynamics of these two cases is the reheating term proportional to $f_-$ in Eq.~(\ref{rec}), which is absent in Eq.~(\ref{adexp}). 
For $T\simeq T_c$ we have $s_+-s_-\simeq L/T_c$ and $s_c\propto \rho_{Rc}$, so this term is proportional to the ratio $L/\rho_{Rc}$.
Thus, Eq.~(\ref{rec}) takes into account the released energy as well as the heat capacity of the plasma. 
Indeed, ignoring the last term, this equation gives a temperature increase  $\Delta T\simeq (dT/d\rho_R) Lf_-$, as expected.

To compute the fraction of volume (either in the detonation or the deflagration case), we shall assume that the new phase is composed of spherical bubbles which may overlap. At time $t$, the radius of a bubble which nucleated at time $t^{\prime}$ is given by 
\begin{equation}
R(t^{\prime},t)=\int_{t^{\prime}}^{t}dt^{\prime\prime}v_{w}(t^{\prime\prime}),\label{radio}
\end{equation}
where we neglected the initial radius. This is generally valid unless $T$ is very close to the Planck scale.
The fraction of volume in
the old phase is given by \cite{gt80,gw81} 
\begin{equation}
f_{+}(t)=\exp\left[-I(t)\right],\label{fmas}
\end{equation}
where 
\begin{equation}
I(t)=\frac{4\pi}{3}\int_{t_{c}}^{t}dt^{\prime}\Gamma(t^{\prime})R\left(t^{\prime},t\right)^{3}.\label{I}
\end{equation}
Here,
$t_{c}$ is the time at which $T=T_{c}$.
In these equations we have ignored the effect of the
scale factor on physical lengths (namely, the stretching of $R$ and the dilution of the density of nucleated bubbles). We shall take into account this effect in the numerical computations, but it is negligible for the cases with little supercooling we consider. Indeed, 
although
for deflagrations the phase transition may last quite longer than for
detonations, we will have a duration of
order $10^{-2}H^{-1}$ and, thus, $a/a_c\simeq1$. Notice, on the other hand, that the
variation of $a$ cannot be ignored in Eqs.~(\ref{adexp}) or (\ref{rec}), since a small change in $T$ will  cause a large change in $\Gamma$.

A measure of progress of the phase transition is given by the fraction of volume  $f_+(t)$.
As in our  work \cite{nos}, we shall define a few reference points in this evolution. The first of them corresponds to the ``initial'' moment $t_{I}$ at which $f_{+}(t_{I})=0.99$. The second one is the percolation
time $t_{P}$, which is approximately given by $f_{+}(t_{P})=0.71$.
Another reference time, which is often considered, is the time $t_{E}$ at which the fraction of volume has fallen to $f_{+}(t_{E})=1/e$. Finally, we define the ``final'' time $t_{F}$ by $f_{+}(t_{F})=0.01$.

\subsection{Numerical results}

\begin{figure}[bt]
\centering
\includegraphics[width=16cm]{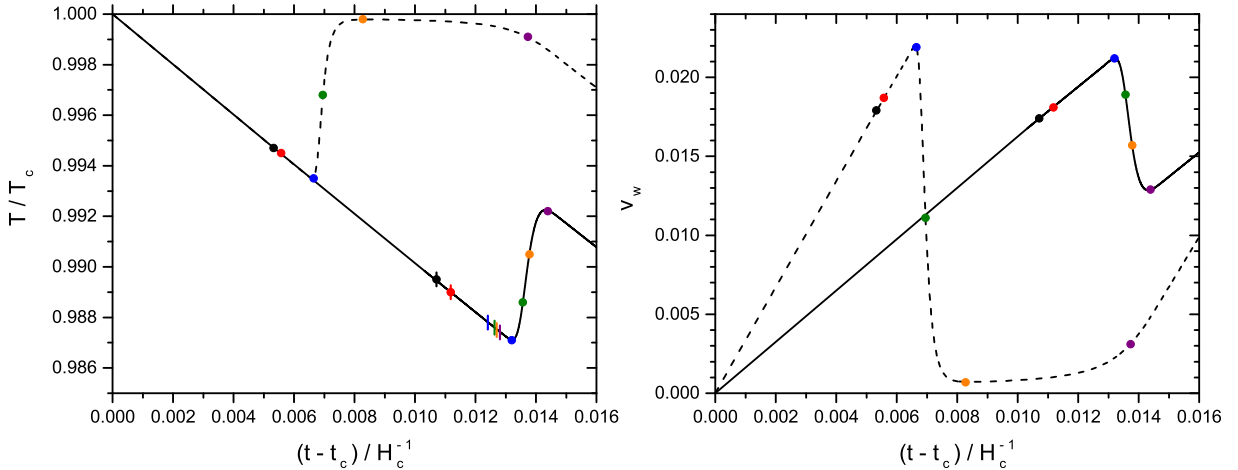}
\caption{Temperature and wall velocity as a function of time, for $L/\rho_{Rc}=0.025$ (solid) and $0.05$ (dashed).
Colored dots and ticks indicate the reference times $t_{H}$ (black), $t_{N}$ (red), $t_{I}$ (blue), $t_{P}$ (green), $t_{E}$ (orange), and $t_{F}$ (purple). The dots correspond to the deflagration case  $\eta/T_{c}^{4}=0.5$ while the ticks correspond to the detonation case $\eta/T_{c}^{4}=0.001$.}
\label{figtempvelo}
\end{figure}

As discussed in previous subsections, we shall fix the potential parameters to typical values, we shall consider $\eta\sim T_c^4$, which gives slow deflagrations, and we shall consider a couple of representative values for the ratio  $L/\rho_{Rc}$. In order to compare with the detonation case, we shall also consider  
the case $\eta\ll T_c^4$.
In the left panel of Fig.~\ref{figtempvelo} we show the evolution of the temperature. 
The time is normalized to the Hubble time $H_c^{-1}$, where $H_c\equiv H(T_c)$. 
The solid curve corresponds to the case  $L/\rho_{Rc}=0.025$. The initial part of the curve corresponds  both to deflagrations ($\eta/T_c^4=0.5$) and detonations ($\eta/T_c^4=0.001$), since 
the supercooling stage does not depend on the friction.
The evolution of the temperature  is initially determined by Eq.~(\ref{adexp}), and the curves would separate once the reheating becomes appreciable in the deflagration case. However, for the detonation the phase transition completes sooner due to the higher wall velocity.

The reference times $t_{H}$, $t_{N}$, $t_I$, $t_P$, $t_E$ $t_F$
are indicated on the curve by dots for the deflagration and by ticks for the detonation. We see that
the first two dots ($t_H$ and $t_N$) coincide with the first two ticks. In contrast,
the rightmost (purple) tick, indicating the time $t_{F}$ for the detonation case, is to the left of the blue dot, which indicates the time $t_{I}$ for the deflagration case. 
This means that, for detonations, a 99\% of space is in the new phase 
even before a 1\% is reached for deflagrations.
In the deflagration case, we observe that, as soon as the fraction
of volume occupied by bubbles begins to be noticeable (i.e., at $t\simeq t_I$) the reheating becomes noticeable too. 
Once bubbles have filled most of space ($t\simeq t_F$, purple dot), the release of latent heat ceases and  the temperature decreases again. The solid line in the right panel corresponds to the wall velocity. We see that the velocity decreases during reheating\footnote{The detonation wall velocity, which is not plotted, varies from $v_{w}\simeq0.95$ at  $t=t_H$ to $v_{w}\simeq0.97$ at $t=t_F$.}, as expected from Eq.~(\ref{vwsimple}).

The dashed curves correspond to the case $L/\rho_{Rc}=0.05$. 
Notice that in this case we have less supercooling that in the previous one. This is because larger $L$ roughly corresponds to a larger parameter $D$, and this gives a smaller value of
$(T_{c}-T_{0})/T_{c}$ [see Eqs.~(\ref{lat}) and (\ref{tct0})], which is reflected in a smaller value of $(T_{c}-T_{N})/T_{c}$. In spite of the smaller supercooling, the maximum velocity is similar to that of the previous case ($v_w\sim10^{-2}$). This is because the pressure difference is roughly proportional to $L$, which is reflected in $v_w$, as can be seen in the approximation (\ref{vwsimple}). On the other hand, 
in this case there are no detonation solutions, no matter how small  the friction coefficient.

Like in the previous case, the temperature begins to grow at $t\simeq t_I$ and decreases again for $t\simeq t_F$. 
In this case, though, this time interval is longer, and we have a stage of approximately constant temperature $T_r$ very close to $T_c$. This occurs due to the larger latent heat, which would actually be enough to reheat the system to a temperature $T>T_c$. Nevertheless,  the backreaction of reheating on  bubble growth prevents this to happen. In our case, 
the released energy gets quickly distributed and the  increase in energy density is given by $Lf_-$, which is initially small.
As the temperature gets close to $T_c$,
the wall velocity decreases significantly, which can be appreciated in the right panel of Fig.~\ref{figtempvelo}. 
The phase transition slows down, preventing further release of latent heat. 
The reheating temperature $T_r$, as well as the wall velocity during this stage, are determined by the balance between the rate at which energy is injected, which is roughly given by $Ldf_-/dt$, and that at which the adiabatic expansion takes energy from the plasma, which is roughly given by $ 4\rho_R H$. This gives the equation $(L/\rho_R)df/dt=H$. We discuss this approximation further in Sec.~\ref{anf}.

These results are in qualitative agreement with previous works (see, e.g., \cite{h95,ma05,m01,ms08}).
The effects of a significant velocity slow-down have been investigated in some detail in 
Refs.~\cite{h95,ma05,m01}. 
In Refs.~\cite{w84,kk86,suhonen,m06}, the limit of a long phase-coexistence stage at $T\simeq T_c$ has been investigated.
Here, we shall consider the general case, and we shall focus on the dynamics of nucleation. 
Some quantities of interest are the average nucleation rate 
$\bar{\Gamma}(t)=f_{+}(t)\Gamma(t)$, 
the number density of bubbles,
\begin{equation}
n(t)=\int_{t_{c}}^{t}dt^{\prime}\bar{\Gamma}(t^{\prime}),\label{n}
\end{equation} 
the average distance between centers of nucleation,
$d(t)=n(t)^{-1/3}$,
the average bubble size, 
\begin{equation}
\bar{R}(t)=n(t)^{-1}\int_{t_{c}}^{t}dt^{\prime}\bar{\Gamma}(t^{\prime})R(t^{\prime},t),
\label{rm}
\end{equation}
and the distribution of bubble sizes,
\begin{equation}
\frac{dn}{dR}(t)=\frac{\bar{\Gamma}(t_{R})}{v_{w}(t_{R})},\label{dndR}
\end{equation}
where $t_{R}$ is the time at which the bubble of radius $R$ was
nucleated, which is obtained by inverting Eq.~(\ref{radio}) for
$t^{\prime}$ as a function of $R$.

\begin{figure}[tbp]
\centering
\includegraphics[width=16cm]{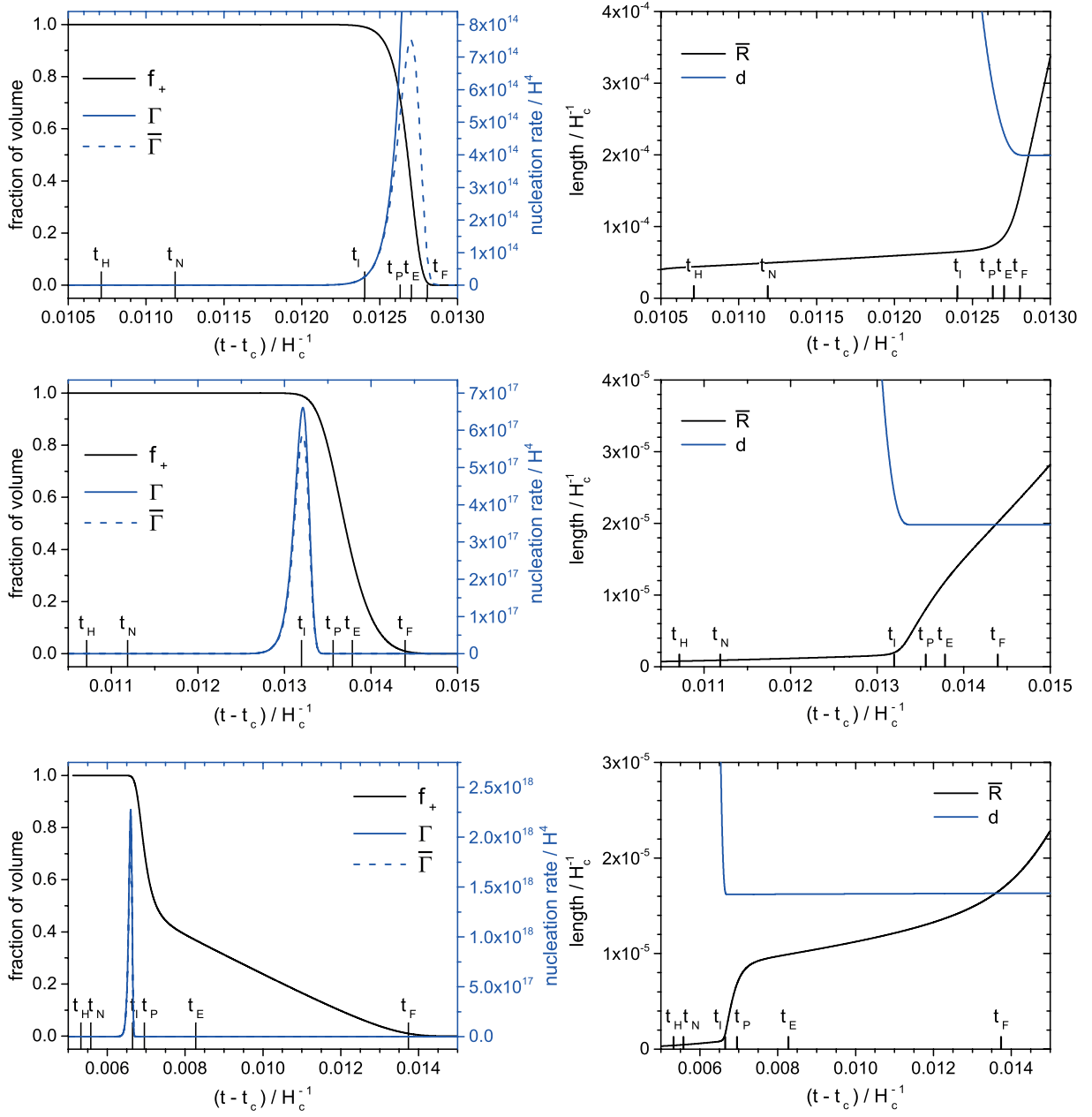}\caption{Evolution of the fraction of volume $f_{+}$, the nucleation rate
$\Gamma$ and its space average $\bar{\Gamma}$ (left panels), and the
average bubble radius $\bar{R}$ and bubble separation $d$ (right
panels). The top row corresponds to the detonation case $L/\rho_{Rc}=0.025$, $\eta=0.001T_{c}^{4}$, 
the central row to the deflagration case $L/\rho_{Rc}=0.025$, $\eta/T_{c}^{4}=0.5$, and the bottom row to the deflagration case
$L/\rho_{Rc}=0.05$, $\eta/T_{c}^{4}=0.5$.}
\label{figevol}
\end{figure}

In Fig.~\ref{figevol} we show the evolution
of some of these quantities for the detonation and the two deflagration cases. 
In the detonation case (upper plots), the development
of the phase transition is 
determined by the extremely quick growth of the nucleation rate (see
the left panel). The main features are the following. Once bubble
nucleation effectively begins, the phase transition completes in a
relatively short time, i.e.,  $t_{F}-t_{N}\ll(t_{N}-t_{c})$.
Moreover, the variation of the fraction of volume occurs in an
even shorter time $t_{F}-t_{I}$.
Most bubbles nucleate in this short interval near $t_F$. In the right panel we see that the
average bubble size grows very slowly during most of the phase transition.
This is due to the constant nucleation of very small bubbles. Only when the
volume in the old phase becomes small and the nucleation of bubbles turns off,
the average bubble radius begins to grow with velocity $v_{w}$.
On the other hand, the average separation between centers of nucleation inherits initially
the rapid variation of $\Gamma$, and becomes constant when  $\bar\Gamma$ vanishes. At $t\simeq t_{F}$ we have $d\sim\bar{R}$, as expected.

The second row of Fig.~\ref{figevol} corresponds to the deflagration
case $L/\rho_{Rc}=0.025$. The model parameters are exactly the same as in the previous case, except for the
friction.
We see that the
evolution of the various quantities is quite different. In particular,
since $\Gamma$ is very sensitive to the temperature, the nucleation
rate turns off as soon as the reheating begins. At this moment we still have $f_{+}\simeq1$,
so $\bar{\Gamma}(t)$ almost coincides with $\Gamma(t)$.
Since bubble nucleation ceases so soon, the average distance $d$ reaches its final value at $t\simeq t_{I}$
(in contrast, in the detonation case this happens at $t\simeq t_{F}$).
Similarly, the transition from an approximately constant average radius
to the behavior $d\bar{R}/dt=v_{w}$ occurs at $t\simeq t_{I}$ and
not at $t\simeq t_{F}$. Notice that in
the deflagration case the final bubble separation is smaller, so the final number of bubbles is higher than in the detonation case. This is because, in this slower phase transition,  
the time at which the nucleation stops is later than  for the detonation case (see the left panels). 
In this small time difference, the nucleation rate reaches quite higher values. 

The effect of reheating is more marked in the third row of plots,
corresponding to the deflagration case with $L/\rho_{Rc}=0.05$. Since
in this case the interval $t_{F}-t_{N}$ is longer (due to the slow-down
after reheating) the nucleation rate looks like a sharp peak at $t\simeq t_{I}$.
Like in the previous case, at $t\simeq t_{I}$ the distance $d$ becomes
constant and the average radius $\bar R$ begins to grow with velocity $v_w$. The subsequent change
of slope of $\bar{R}$ and $f_{+}$ corresponds to the decrease of $v_{w}$. 

\begin{figure}[bt]
\centering
\includegraphics[width=16cm]{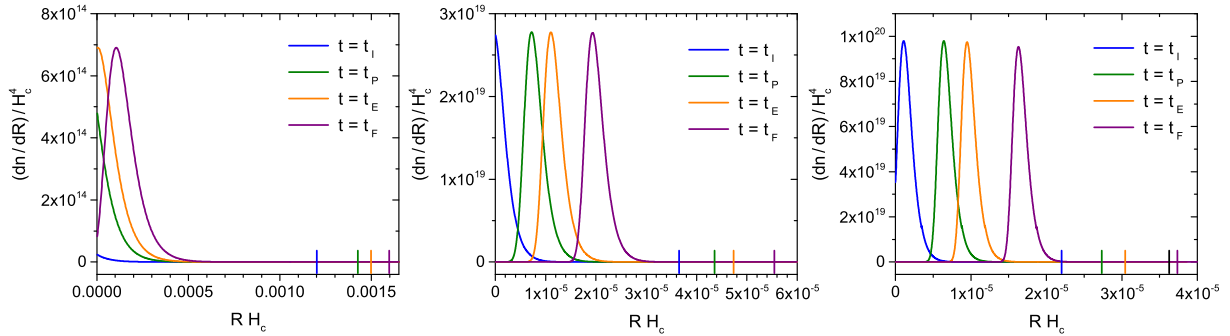}
\caption{The bubble size distribution at the times $t_{I}$, $t_{P},$ $t_{E}$,
$t_{F}$, for the case $L/\rho_{Rc}=0.025$, $\eta=0.001T_{c}^{4}$
(left), $L/\rho_{Rc}=0.025$, $\eta/T_{c}^{4}=0.5$ (center), and $L/\rho_{Rc}=0.05$,
$\eta/T_{c}^{4}=0.5$ (right). The ticks indicate the value of $R(t_{N},t)$
at each of these times. \label{figdistrib}}
\end{figure}

In Fig.~\ref{figdistrib} we plot the distribution of bubble sizes
for the three cases at the times $t_{I}$, $t_{P}$, $t_{E}$ and $t_{F}$.
We also indicate, for comparison, the size (at each of these times)
of a bubble which nucleated at $t=t_{N}$. 
In the fast wall case, we see that the size distribution maximizes at $R=0$ until the phase transition is quite advanced (more precisely, until $f_{+}=e^{-1}$). This is again due to
the rapidly increasing nucleation of vanishingly small bubbles until the time $t_E$.
In contrast, in the slow wall case,  for $t\gtrsim t_{I}$ no new bubbles are nucleated, and the size distribution just shifts to larger radius. Hence, 
the maximum separates from $R=0$. 
For the same reason, the ratio of the average radius $\bar{R}(t)$
to that of the largest bubbles, $R(t_{N},t)$ is smaller in the detonation
case than in the deflagration case. 

\begin{figure}[bt]
\centering
\includegraphics[width=16cm]{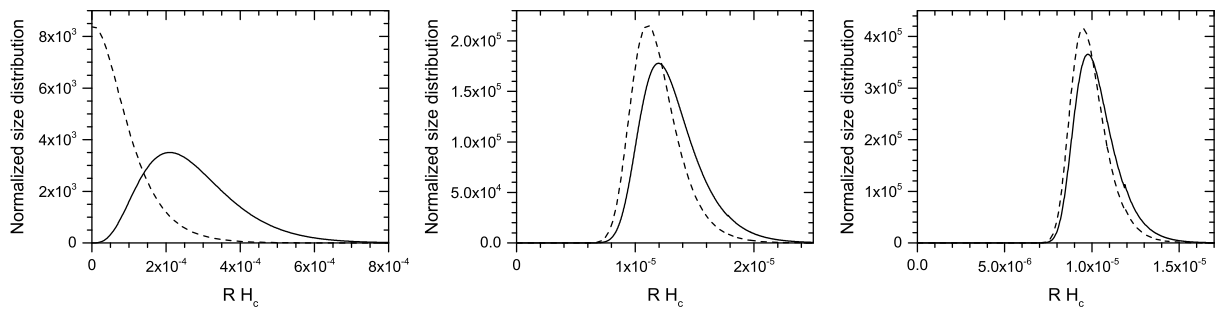}
\caption{The bubble size distributions $(R^3dn/dR)/(\int R^3dn)$ (solid lines)
and $(dn/dR)/n$ (dashed lines)
at $t=t_E$ for the three cases of Fig.~\ref{figdistrib}.} \label{figdistrcub}
\end{figure}

For some applications, such as gravitational-wave generation, it is more appropriate to consider the volume-weighted average radius. However, in the case of slow deflagrations, there will  not be a large difference between the weighted and unweighted averages, since all bubbles nucleate around the same time $t\simeq t_I$ and, hence, have similar sizes.  To illustrate this, in Fig.~\ref{figdistrcub} we show the radius distribution together with the volume-weighted distribution at the time $t=t_E$. We see that only in the detonation case the difference may be relevant.

\section{Approximations for the nucleation rate \label{approxrate}}

In the computations of the previous section we have made several approximations in order to simplify the treatment. However, much simpler approximations are often used, such as assuming a   
constant wall velocity and even a constant nucleation rate. As we have seen, none of these is a good approximation in our case. For the wall velocity we may use the relatively simple form (\ref{vwsimple}). On the other hand, 
for the nucleation rate, it is not easy to find a simple approximation.
For $T$ close to $T_c$, the thin wall approximation can be used
for the instanton action, and we have $S(T)\propto1/V(\phi_{-},T)^{2}$
\cite{ah92}. Linearizing $V$ as in Eq.~(\ref{fmelin}), we obtain a nucleation rate of the form
\begin{equation}
\Gamma=A\exp[B/(T_{c}-T)^{2}],\label{nucanal}
\end{equation}
where the constants $A$ and $B$ depend on the potential parameters $D,E,\lambda$ (see, e.g., \cite{m04}). This expression shows that $\Gamma$ is a very rapidly varying function of $T$, and assuming $\Gamma=$ constant will generally be a bad approximation. From Eq.~(\ref{nucanal}) we may obtain $\Gamma(t)$ using the appropriate time-temperature relation, such as Eq.~(\ref{adexp}) or Eq.~(\ref{rec}).
However,  analytic approximations for the nucleation rate may  introduce large errors.
Therefore, it is usual to consider a semi-analytic approach, which consists in 
linearizing the exponent $S(t)$ in Eq.~(\ref{gamma}) around a certain time $t_*$. This procedure only requires to compute numerically $S$ and its derivative at $T_*=T(t_*)$.

\subsection{Exponential rate}

As pointed out in Ref.~\cite{nos}, the linearization of $S(t)$ actually involves linearizing both functions $S(T)$ and $T(t)$, and any of these approximations may fail.
Nevertheless, except in special cases of very strong phase transitions, the temperature $T_*$ will not be close to a minimum of $S(T)$, and we may expand $S$ to first order,
\begin{equation}
S(T)\simeq S(T_{*})+(dS/dT)|_{T_{*}}(T-T_{*}), \label{Slin}
\end{equation}
provided that the temperature variation is small
enough. 
Assuming that (in the radiation-dominated era) the temperature decreases
with time as $dT/dt=-HT$, then for a short enough time we may write
\begin{equation}
T-T_{*}=-H_*T_{*}(t-t_{*}),  \label{Ttlin}
\end{equation} 
with $H_*=H(T_*)$. Hence, we have an exponentially growing nucleation rate 
\begin{equation}
\Gamma(t)=\Gamma_{*}\exp[\beta_*(t-t_{*})],\label{exp}
\end{equation}
where $\Gamma_{*}=\Gamma(T_{*})$ and  $\beta_*=(HTdS/dT)|_{T_{*}}$.  
This rate is simpler than Eq.~(\ref{nucanal}), and may be a much better approximation than the latter if $t_*$ is conveniently chosen, so that the values of interest of $t$ are close to it\footnote{The quadratic correction to this linearization has been considered recently in Ref.~\cite{jst17}.}.
The exponential rate is a good approximation
for a phase transition mediated by detonations (see \cite{nos} for a recent discussion). 
On the other hand, as can be appreciated in  Fig.~\ref{figevol},
this is not a good approximation for deflagrations. This is because  Eq.~(\ref{Ttlin}) will no longer be valid in the presence of reheating.

Nevertheless, this approximation will always be  valid in the initial stages of the phase transition.
For the initial number density (\ref{tN}), the nucleation rate (\ref{exp}) gives  $n(t) \simeq\Gamma(t)/\beta_*$, and the condition for the onset of nucleation is 
\begin{equation}
\Gamma(t_N)/\beta_*=H(t_N)^3 \label{eqtN}
\end{equation}
The quantity $\beta_*$ is quite sensitive to the temperature, which implies that the exponential rate is a good approximation only in a small interval around $t_*$. Nevertheless, since $\Gamma$ grows quickly with time, the interval of interest is generally small.
For instance, in Eq.~(\ref{tN}) it is convenient to choose  $t_*=t_N$, since $\Gamma$ decreases rapidly for smaller times. Thus, 
Eq.~(\ref{eqtN}) gives the equation
\begin{equation}
S(T_{N})-(3/2)\log[S(T_{N})/(2\pi)]+\log[T_{N}S^{\prime}(T_{N})]=4\log\left[T_{c}/H_{c}\right]
\label{TNap}
\end{equation}
(in the last term we have used the approximation $T_{N}\simeq T_{c}$). 
This equation gives a very good estimate for the temperature $T_N$  \cite{nos}.

For the detonation case, we may use the exponential rate beyond $t=t_N$.
For a small time interval we may also assume a constant wall velocity, in which case Eqs.~(\ref{radio}-\ref{I}) can be integrated analytically \cite{tww92}. We have 
\begin{equation}
I(t)=8\pi v_{w}^{3}\Gamma(t)/\beta_*^{4}.\label{Iexp}
\end{equation}
From Eq.~(\ref{Iexp}) we may obtain analytic expressions for quantities such as
$n$ or $dn/dR$, as well as analytic estimations of
time intervals.  
These analytic results  can be applied to physical
models by computing the parameters $\Gamma_{\ast}$, $\beta_*$ and
$v_{w}(T_{\ast})$. Moreover, the parameter $\Gamma_{\ast}$ is not too relevant,
since the dynamics of nucleation depends essentially on  $\dot \Gamma/\Gamma\simeq\beta_*$. The parameter $\beta_*^{-1}$ is the time scale for the dynamics.
From Eq.~(\ref{TNap}), we have $S\sim4\log(M_{P}/v)\gg1$, and one may expect similar values for  $\beta_*/H=TdS/dT$.
Actually, for $T_*$ close to $T_{c}$ we have even higher
values, since Eq.~(\ref{nucanal}) gives $dS/dT=2S(T)/(T_{c}-T)$. Hence, 
we have $\beta_*/H\sim8\log(M_{P}/v)(1-T_*/T_{c})^{-1}$. 
For our numerical examples we have $S\sim10^{2}$ and $\beta_*/H\sim10^{4}$.
Thus, in the detonation case we obtain, e.g., $t_{F}-t_{I}\sim\beta_*^{-1}\sim10^{-4}H^{-1}$, in agreement with Fig.~\ref{figevol}. 
On the other hand, the time elapsed since the critical temperature is given by  $H(t_*-t_{c})\simeq1-T_*/T_{c}$,
so we have $\beta_*^{-1}/(t_*-t_{c})\sim [8\log(M_{P}/v)]^{-1}$. Therefore,
the time $\beta_*^{-1}$  characterizing the dynamics will be generally
much smaller than $t_*-t_c$. This is also in agreement with the top panels of Fig.~\ref{figevol}.

In the deflagration case, these results will  be valid as long as the reheating is not appreciable. 
According to Eq.~(\ref{rec}), the reheating rate will be roughly proportional to $df_-/dt$.
From Eqs.~(\ref{eqtN}-\ref{Iexp}), at $t=t_N$ we have $I(t_{N})\simeq8\pi v_{w}^{3}(H/\beta_*)^{3}\ll 1$.
For instance, in our numerical examples  ($v_{w}\sim10^{-2}$, $\beta_*/H\sim10^{4}$),
we have $I(t_{N})\sim10^{-17}$. Therefore, at this stage we have $f_{-}\simeq I\ll1$ and $df_-/dt\simeq\beta_* I\ll H$. Hence, according to Eq.~(\ref{rec}), at $t=t_N$ we certainly still have adiabatic cooling, and the supercooling will continue for a considerable time, in agreement with Fig.~\ref{figtempvelo}.

\subsection{Sudden reheating}

For $t>t_N$, the reheating will be noticeable at some point, and the exponential rate (\ref{exp}) will break down.  
Since the temperature variations are relatively small, we may still use the linear approximation (\ref{Slin}) for $S(T)$. 
However, we must replace the linear time-temperature relation
(\ref{Ttlin}) with the relation (\ref{rec}), which takes into account the reheating. In order to obtain analytic results, we need to simplify further this equation. We shall accomplish this by assuming 
a completely homogeneous temperature, $T_{-}\simeq T_{+}\equiv T$. This is valid since $T_+-T_-\sim (L/\rho_{Rc})(T_c-T_+)$ (see the appendix).
Using this approximation in Eq.~(\ref{rec}), we obtain the simple relation\footnote{This simpler approximation has already been considered in Ref.~\cite{m04}. Since hydrodynamics is neglected, the bag approximation is not necessary. We have $\Delta s=\partial V/\partial T$ and $s_{c}=2\pi^{2}g_{*}T_{c}^{3}/45$.}
\begin{equation}
T^{3}/T_{c}^{3}=(\Delta s/s_{c})f_-+a_c^3/a^{3},\label{rechom}
\end{equation}
where $\Delta s(T)=s_+(T)-s_-(T)$. If we differentiate Eq.~(\ref{rechom}), to lowest order in $T-T_c$ and $L/\rho_{Rc}$  we obtain  (see the appendix for details)
\begin{equation}
3\dot{T}/T_{c}=r\dot{f}_{-}-3H_c,\label{tTsimple}
\end{equation}
where a dot indicates a derivative with respect to time, and we have defined the parameter
\begin{equation}
r\equiv\frac{\Delta s(T_c)}{s(T_{c})}=\frac{3}{4}\frac{L}{\rho_{Rc}}.
\end{equation}
In terms of bag parameters, we have $r=3\alpha(T_c)$.

As we have seen, for $t\simeq t_N$ we have $\dot f_-\simeq 0$. In the approximation (\ref{tTsimple}), at this stage the temperature decreases linearly,  $\dot{T}\simeq-T_cH_c$. This behavior is observed in Fig.~\ref{figtempvelo}. In the units of this figure, the slope of the curve is $-1$.
In our numerical examples, the reheating becomes noticeable  for $t\simeq t_{I}$, i.e.,  for $f_-\simeq I \sim10^{-2}$. 
In the general case, this will happen when the reheating term in Eq.~(\ref{tTsimple}),  $r\dot{f}_{-}$, becomes comparable to the adiabatic cooling term $3H_c$. 
Assuming that this happens for small $f_-$, we have $f_{-}\simeq I$. Assuming also that the dynamics is still characterized by the time $\beta_*^{-1}$, we have $\dot{f}_{-}\sim\beta_* I$. We thus obtain the condition $f_{-}\sim(3/r)(H_c/\beta_*)$ for the reheating to become noticeable. Notice that, even though $r$ is a small number, $H_c/\beta_*$ is even smaller, so the approximation $I\ll1$ is generally consistent. For our example cases, this estimation gives $f_{-}\sim10^{-2}$. 

More precisely, the equality of the cooling and reheating terms in  Eq.~(\ref{tTsimple}),
\begin{equation}
r\dot f_-(t_m)=3H_c,  \label{eqTmin}
\end{equation} 
gives the condition for the minimum temperature $T_m$.
In order to estimate the time $t_m$, we notice that  $\dot{f}_{-}(t)$ has an extremely rapid growth at this stage, so  an instant before $t_m$ the reheating term  in (\ref{tTsimple}) was negligible. Therefore, we have adiabatic cooling  until almost $t=t_m$. This can be appreciated in Fig.~\ref{figtempvelo}. Then, to solve the condition (\ref{eqTmin}) 
we may use the exponential-rate approximation. Assuming small $I$, we have $\dot f_-\simeq \dot I$, and using the result (\ref{Iexp}) we have
$8\pi v_{w}^{3}\Gamma(t_m)/\beta_*^{3}=3H_c/r$. We may also choose $t_*\simeq t_m$, and we obtain
\begin{equation}
\Gamma_{m}=\frac{3H_c\beta_m^{3}}{8\pi rv_{m}^{3}}, \label{gamtrunc}
\end{equation}
where $\Gamma_m=\Gamma(T_m)$, $v_m=v_w(t_m)$, and
\begin{equation}
\beta_m\equiv(HTdS/dT)_{T_m}. \label{betam}  
\end{equation}
This gives a semi-analytic equation for $T_m$,
\begin{equation}
S(T_{m})-\frac{3}{2}\log S(T_{m})+3\log[T_{m}S^{\prime}(T_{m})]=4\log\left[T_{c}/H_{c}\right]+\log(\sqrt{{8}/{\pi}}\,{r}\,v_{m}^{3}/3).\label{Sm}
\end{equation}
The wall velocity in the last term can be replaced by the estimation (\ref{vwsimple}) as a function of $T_m$. Moreover, due to the logarithmic dependence, it can be even replaced by the constant $v_{w}(T_{N})$
with no significant error. With the latter simplification, the approximation (\ref{Sm}) gives the value of $T_{m}$ with a relative error of order $10^{-4}$ with respect to the numerical
computations of Sec.~\ref{dyn}. Using the result in the linear approximation 
(\ref{Ttlin}) (with $t_*=t_m$), we obtain $t_m-t_c$ with a relative error of order $10^{-3}$.

This estimation of $t_m$ is so good because, in the first place, the reheating is almost instantaneous, which justifies taking the limit $t_*\to t_m$. In the second place, the use of Eq.~(\ref{Iexp}) involves a time interval of order $\beta_m^{-1}$. 
We may obtain estimations for longer time intervals using this approximation; however, the errors will be larger.
For instance, the time $t_m-t_H$ is obtained by 
comparing the result (\ref{gamtrunc}) with $\Gamma(t_{H})=H^{4}$. This gives  
\begin{equation}
t_{m}-t_{H}\simeq\beta_m^{-1}\log\left[\frac{3}{8\pi r}\left(\frac{\beta_m}{H_cv_{m}}\right)^{3}\right].
\label{tHtm}
\end{equation}
Similarly, comparing (\ref{gamtrunc}) with $\Gamma(t_{N})\simeq\beta_m H^{3}$, we obtain
\begin{equation}
t_{m}-t_{N}\simeq\beta_m^{-1}\log\left[\frac{3}{8\pi rv_{m}}\left(\frac{\beta_m}{H_cv_{m}}\right)^{2}\right], \label{tNtm}
\end{equation}
These analytic relations 
assume an exponential rate, as well as a constant wall velocity, for  time intervals which are several times $\beta_m^{-1}$ (e.g., $t_m-t_H\sim 10\beta_m^{-1}$). For our examples, the agreement  with the numerical computation is  around a 25\%.  

Since $f_-$  grows so rapidly at $t=t_m$, we may assume  a sudden reheating at this point, such that $\Gamma(t)$ vanishes for $t>t_m$. This corresponds to a nucleation rate of the form
\begin{equation}
\Gamma(t)=\begin{cases}
\Gamma_{m}\exp[\beta_m(t-t_{m})] & \mbox{ for } t\leq t_{m},\\
0 & \mbox{ for }  t>t_{m}.
\end{cases}\label{gammatrunc}
\end{equation}
Since we have $f_{+}\simeq1$ for $t<t_m$, we can make the replacement $\bar{\Gamma}(t)\simeq\Gamma(t)$ in  Eqs.~(\ref{n}-\ref{dndR}), which simplifies the computation of the quantities $n$, $\bar{R}$ and $dn/dR$.
Moreover, the final number density of bubbles is given by $n_f=n(t_m)$, and we have
\begin{equation}
n_f=\Gamma_m/\beta_m,\label{niexp}
\end{equation}
so the final average distance between centers of nucleation is given by
\begin{equation}
d_f=\left(\frac{8\pi r\beta_{m}}{3H_c}\right)^{1/3}v_{m}\beta_{m}^{-1}.\label{dexp}
\end{equation}
This estimation gives the value
of $d_f$ to a $6\%$ of the numerical computation. 
On the other hand, assuming a constant wall velocity for $t<t_m$, the exponential rate gives a constant value for 
the average bubble radius, 
\begin{equation}
\bar{R}(t)=v_{m}\beta_m^{-1}.\label{rmexp}
\end{equation}
This is in agreement with the initial behavior of $\bar{R}$ in the plots on the right of Fig.~\ref{figevol}. The small slope in the plots
is due to the fact that $v_{w}$ is not constant and the nucleation rate is not exactly exponential. For $t>t_m$, the computation of $\bar R(t)$  is involved even with these simplifications, and we shall consider a different approximation for $\Gamma(t)$ below. Nevertheless, notice that the value of $\bar R$ at $t=t_F$ is approximately given by the average bubble separation, $\bar{R}(t_F)\simeq d_f$. In our examples, the value (\ref{dexp}) is  a factor $\sim10$ larger than the  initial value (\ref{rmexp}).

\subsection{Gaussian nucleation rate}

The approximation (\ref{gammatrunc}) gives a function $\Gamma(t)$
with a sharp peak at $t=t_{m}$, while  the actual nucleation
rate has a differentiable maximum. 
As we have seen, for the computation of several quantities this qualitative difference is not important. However, it will be reflected, for instance, in the shape of the bubble size distribution. 
To obtain a smooth function, we may use the quadratic expansion of $T(t)$ around its minimum  at $t_m$. Together with 
the linear approximation (\ref{Slin}) for $S(T)$ around $T_*=T_m$, this gives
a Gaussian nucleation rate, 
\begin{equation}
\Gamma(t)=\Gamma_{m}\exp[-\alpha_m^{2}(t-t_{m})^{2}],\label{gauss}
\end{equation}
where 
\begin{equation}
\alpha_m^{2}=\left.\frac{dS}{dT}\right|_{T_m}\frac{1}{2}\left.\frac{d^{2}T}{dt^{2}}\right|_{t_{m}}.
\label{alfa0}
\end{equation}
This approximation will be valid only for $t$ close enough to $t_{m}$.
Nevertheless, away from $t=t_{m}$ the nucleation rate decreases rapidly
and its value is less relevant. 

The first factor in Eq.~(\ref{alfa0}) is related to the constant $\beta_m$  defined in Eq.~(\ref{betam})\footnote{
Notice that $\beta_m$
is defined as a function of the temperature $T_{m}$ and does not
depend on the dynamics. In particular, 
$\beta_m$ does not coincide with $dS/dt|_{t_{m}}$, as it would in the absence of reheating.
In the present case the latter derivative vanishes, since $S$ has a minimum at $t_m$.}.
On the other hand, from
Eq.~(\ref{tTsimple}) we obtain the last factor,
$\ddot{T}=rT_{c}\ddot{f}_{-}/3$.  
For $I\ll1$ we have $\ddot{f}_{-}=\ddot{I}$, and we may write 
\begin{equation}
\alpha_m^{2}=\frac{1}{2}\frac{\beta_{m}}{H}\frac{r}{3}\ddot{I}(t_{m}),\label{alfagral}
\end{equation}
We have seen that  Eq.~(\ref{Sm}) provides a good estimation for the values of $T_{m}$ and $t_m$.
To compute $\ddot{I}(t_{m})$
from Eqs.~(\ref{radio}-\ref{I}), we must take into account that 
the velocity has a maximum at $t=t_m$, as can be seen either from Eq.~(\ref{vwnr}) or  Eq.~(\ref{vwsimple}), and also in Fig.~\ref{figtempvelo}. We thus have 
\begin{equation}
\ddot{I}(t_{m})=8\pi v_{m}^{2}\int_{t_{c}}^{t_{m}}dt\Gamma(t)R(t,t_{m}).\label{Iseg}
\end{equation}
The right hand side does not involve any derivatives at $t=t_m$, so we can use again the approximation (\ref{gammatrunc}) to estimate the integral. Notice that the latter is given
by $n(t_{m})\bar{R}(t_{m})=\Gamma_m v_m/\beta_m^2$. Finally, using Eq.~(\ref{gamtrunc}), we obtain 
\begin{equation}
\alpha_m=\frac{1}{\sqrt{2}}\beta_{m}.\label{alfatrunc}
\end{equation}
The characteristic time for the dynamics of bubble nucleation is thus $\alpha_m^{-1}\sim\beta_m^{-1}$.
For the detonation case, the time $\beta_*^{-1}$ roughly characterizes the duration
of the whole phase transition. In contrast, for the deflagration case,  $\beta_{m}^{-1}$
only gives the time in which bubble nucleation occurs. The transition is longer, due to the decrease of both the nucleation rate and the wall velocity.  

For a Gaussian nucleation rate and a constant wall velocity, the evolution of $I(t)$ has been obtained analytically in Ref.~\cite{nos} (for a very strong phase transition with $v_{w}\simeq1$).
In the present case, the approximation $v_{w}\simeq v_{m}$
is valid only for $t\simeq t_{m}$. 
We shall not be interested in the evolution of $f_-(t)$ during a small interval in which $f_-\simeq 0$. 
Nevertheless, in this short time the bubble size distribution is formed. In Eq.~(\ref{dndR}),
$t_{R}$
is the nucleation time of a bubble which has radius $R$ at time $t$. Since most bubbles nucleate around $t=t_m$ (where  $f_+\simeq 1$), we have $\bar{\Gamma}(t_R)\simeq\Gamma(t_R)$ and $v_w(t_R)\simeq v_m$. 
Now, we must invert the relation $R(t',t)$ to obtain $t'=t_{R}(t)$, but 
we are only interested in times $t'$ within a short interval around $t_{m}$. Therefore, we may use the approximation  $R(t',t)\simeq v_{m}(t_{m}-t')+R(t_{m},t)$, in which
all the bubble sizes have approximately the same evolution $R(t_m,t)$, except for the initial dispersion, $v_{m}(t_{m}-t')$.
Inverting this relation, we obtain $t_{R}=t_{m}-[R-R(t_m,t)]/v_{m}$.
For any of our approximations (\ref{gammatrunc}) or (\ref{gauss}),
we see that $dn/dR$ will be a function of $t_{R}-t_{m}=-[R-R(t_m,t)]/v_{m}$. For the Gaussian case we have
\begin{equation}
\frac{dn}{dR}(t)=\frac{\Gamma_{m}}{v_{m}}e^{-\left(\frac{\alpha_m}{v_{m}}\right)^{2}\left[R-R(t_m,t)\right]^{2}},\label{dndRap}
\end{equation}
and we see that for this distribution we have $\bar{R}(t)=R(t_m,t)$. For the distribution given by Eq.~(\ref{gammatrunc}), the latter equality will be a good approximation at late times, due to the relatively small dispersion $\Delta R\sim v_m\beta_m^{-1}$.

\begin{figure}
\centering
\includegraphics[width=16cm]{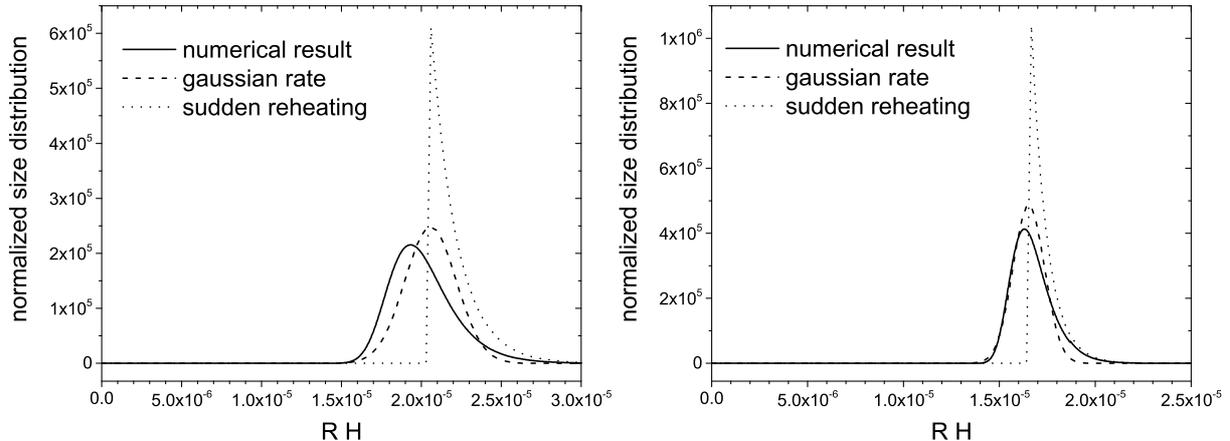}\caption{The normalized size distribution 
$n^{-1}dn/dR$ at $t=t_F$, for the numerical computation (solid lines), the Gaussian approximation
(\ref{gdist})  (dashed lines), and the result obtained for the  rate  (\ref{gammatrunc}) (dotted lines). 
The left panel corresponds to the case $L/\rho_{Rc}=0.025$ and the right panel to the case $L/\rho_{Rc}=0.05$.}
\label{figdistribaps}
\end{figure}

If a particular moment in the development of the phase transition
is characterized by the value of $\bar{R}$, we may evaluate this
distribution without even solving the evolution. 
For instance, at $t=t_F$ we have $\bar{R}\simeq d$ (below we quantify the error of this approximation), where  $d$ takes its final value $d_f$
given by Eq.~(\ref{dexp}). Thus, we have a fully analytic approximation for the size distribution at the end of the transition,
\begin{equation}
\frac{1}{n}\frac{dn}{dR}(t_{F})\simeq\frac{\beta_{m}}{\sqrt{2\pi}v_{m}}\exp\left[-\left(\frac{\beta_{m}}{\sqrt{2}v_{m}}\right)^{2}(R-d_f)^{2}\right].\label{gdist}
\end{equation}
A similar expression is obtained for the sudden-reheating approximation.
In Fig.~\ref{figdistribaps} we compare these approximations and
the numerical computation. 
We consider the normalized distribution $n^{-1}dn/dR$ for a better comparison (so that order 1 errors in the analytic determination
of $\Gamma_{m}$ cancel out). 
We see that the Gaussian nucleation rate gives a better approximation. We remark, though, that in both analytic approximations the position of the peak, $R=d_f$, was computed from the sudden-reheating result (\ref{dexp}).  

\subsection{Delta function rate}

The evolution of the phase transition for $t>t_m$ is difficult to describe analytically. The integrals (\ref{radio})-(\ref{fmas}), involving $R(t',t)$ and $v_w(t'')$, are not easy to solve, except in the simple case of a constant $v_w$, which gives  $R(t',t)=v_w (t-t')$.
Even for the simple approximation (\ref{vwsimple}) for $v_w(T)$  and using the simple relation (\ref{tTsimple}) between $T$, $t$ and $f_-$,  we have an integro-differential equation for $R$ and $v_w$.  The main problem is that we have to deal, even after $\Gamma$ turns off, with bubbles which nucleated at different times $t'$.
Therefore, 
a considerable simplification is achieved by assuming that all bubbles nucleate at $t=t_m$
(we partially used this approximation in the previous subsection).
In this case, all bubbles have the same radius, and we deal with a single variable
$R(t_{m},t)$. 
Thus, we consider a nucleation rate of the form 
\begin{equation}
\Gamma(t)=n_{f}\,\delta(t-t_{m}), \label{gdelta}
\end{equation}
where $n_f$ is the final number density of bubbles, which can be estimated from Eq.~(\ref{niexp}). Integrals like (\ref{rm}) and (\ref{I}) are now trivial. We have $\bar{R}(t)=0$ for $t<t_m$ and $\bar{R}(t)=R(t_{m},t)$ for $t>t_m$.
Thus, we may obtain a relatively simple equation for $\bar{R}(t)$ since we have
\begin{equation}
d\bar{R}/dt=v_{w}(T) \label{vwR}
\end{equation}
for $t>t_m$,
where the temperature depends on $t$ and $f_-$. 
We have $I=(4\pi/3)n_{f}\bar{R}^{3}$, so
\begin{equation}
f_+=\exp[-({4\pi}/{3})\left(\bar{R}/d_f\right)^3]
\label{fmaR}
\end{equation}
(with $d_f=n_f^{-1/3}$). 

Even without solving the equation for $\bar R$ (which we do in the next section), we may check that this approximation is suitable to estimate quantities at later times. 
According to Eq.~(\ref{fmaR}), the equality
$\bar R=d_f$ occurs for $f_{+}=\exp(-4\pi/3)\simeq0.015$. This happens before the time $t_F$, which is defined by the condition $f_+=0.01$. In this approximation, the latter corresponds to $\bar{R}\simeq 1.032d_f$. 
Thus, we see that   
the equality $\bar{R}=d$ occurs very close  to the time $t_F$. This is in agreement with the numerical computations of Sec.~\ref{dyn} and justifies the approximation $\bar{R}(t_F)\simeq d_f$ in Eq.~(\ref{gdist}).

\section{Analytic calculation of the evolution} 
\label{anf}

We shall now compute the evolution of the phase transition
for $t>t_m$ with the approximation of a delta-function nucleation rate. For this aim we consider Eq.~(\ref{vwR}) with
the approximation (\ref{vwsimple}) for the function $v_w(T)$, which we write in the form
\begin{equation}
v_{w}=v_{m}\,\frac{T_c-T}{T_{c}-T_{m}}. \label{vwvm}
\end{equation}
On the other hand, in the present approximation, the linearized equation (\ref{tTsimple}) gives the  time-temperature relations $T_c-T_m= T_cH_c(t_m-t_c)$ and
\begin{equation}
\frac{T-T_{m}}{T_{c}-T_{m}}=qf_{-}-\frac{t-t_m}{t_m-t_c}, \label{Tqtau}
\end{equation} 
where we have defined the quantity
\begin{equation}
q\equiv\frac{r/3}{1-T_{m}/T_{c}}, \label{q}
\end{equation}
which parametrizes the reheating. 
Thus, we have
\begin{equation}
\frac{v_{w}}{v_{m}}=1-qf_{-}+\frac{t-t_m}{t_m-t_c}. \label{vwqtau}
\end{equation}
From Eqs.~(\ref{vwR}), (\ref{fmaR}) and (\ref{vwqtau}), the equation for $\bar R$ becomes 
\begin{equation}
\frac{d\bar{R}}{dt}=
\begin{cases}
0 &  \mbox{ for }  t<t_{m},\\
v_m \left[\frac{t-t_{c}}{t_m-t_c}-q\left(1-e^{-\frac{4\pi}{3}\left({\bar{R}}/{d_f}\right)^{3}}\right)\right] &  \mbox{ for }  t\geq t_{m}.
\label{eqR}
\end{cases}
\end{equation}
By means of Eq.~(\ref{fmaR}), this equation can be converted into an equation for $f_{\pm}$ or for $I$.

\subsection{General behavior}

From Eq.~(\ref{eqR}) we see that the evolution of the phase transition depends on a few parameters, namely, the wall velocity $v_m$, the time $t_m-t_c$, the bubble separation $d_f$, and the parameter $q$.
Under the present approximations, 
this parameter gives the ratio of the released
energy to the energy which is needed to reheat the system from $T=T_m$ back to $T=T_{c}$,
\begin{equation}
q\simeq \frac{L}{\rho_{R}(T_{c})-\rho_{R}(T_{m})}. \label{qratio}
\end{equation}
Therefore, we may expect qualitative differences for $q<1$ and $q>1$. Indeed, in our two numerical examples, we have $q\simeq0.48$ for the case $r=0.025$ and $q\simeq1.89$ for the case $r=0.05$, and the difference is clear in Figs.~\ref{figtempvelo} and \ref{figevol}.

For $q<1$, we see from Eq.~(\ref{Tqtau}) that the temperature reached during reheating is bounded by $T<T_m+q(T_c-T_m)$. 
This maximum value can only be reached if the second term on the right-hand side of Eq.~(\ref{Tqtau}) is negligible, for which the variation of $f_-$ must be very rapid.
Initially, this is the case.  For $t$ close enough to $t_m$, 
Eqs.~(\ref{vwR}) and (\ref{fmaR}) give
\begin{equation}
f_+\simeq \exp\left[{-\frac{4\pi}{3}\left(\frac{v_{m}(t-t_{m})}{d_f}\right)^{3}}\right],
\label{fmaap}
\end{equation}
which has a variation of order 1 in a time $t-t_m\sim d_f/v_{m}$.
From our previous results, this is typically $\sim 10\beta_m^{-1}\ll t_m-t_c$, so the last term in Eq.~(\ref{Tqtau}) can indeed be neglected during a time of this order.
The approximation (\ref{fmaap}) will break down only
if $v_w$ decreases significantly from its maximum $v_m$. 
According to Eq.~(\ref{vwqtau}),
the wall velocity will decrease at most by a factor $1-q$. 
Except in the limit $q\simeq1$, this will be an order 1 factor, and the variation of $f_+$ will occur in a time $t_F-t_m\sim d_f/v_{m}\ll t_m-t_c$. 
At $t=t_F$, the temperature will reach the maximum 
$T_{r}\simeq T_m+q(T_c-T_m)\simeq T_{m}+(r/3)T_{c}$, and
the wall velocity will reach a minimum $v_{r}\simeq v_{m}(1-q)$. 
For $t>t_F$, the temperature is given by $T\simeq T_r-T_cH_c(t-t_{m})$.

On the other hand, for the case $q>1$, neglecting the term proportional to $t-t_m$ in (\ref{Tqtau}) gives a reheating $T_r>T_c$. Nevertheless, this term cannot be neglected in this case, since, as $T$ gets close to $T_c$, the wall velocity (\ref{vwvm}) decreases significantly, and the growth of $f_-$ slows down.
Although we will still have a variation $f_-\sim 1$ in a time of order $d_f/v_{m}$, 
according to Eq.~(\ref{vwqtau}), when the fraction of volume
reaches a value $f_{-}\simeq1/q$ we will have $v_{w}/v_m\ll 1$. At this point the phase transition enters 
a phase-equilibrium stage at $T\simeq T_{c}$. 
In this stage we have,  from Eq.~(\ref{Tqtau}), the  evolution
\begin{equation}
f_{-}\simeq \frac{1}{q}\left(1+\frac{t-t_m}{t_m-t_c}\right). \label{fslow}
\end{equation}
Hence, the condition $f_{-}=1$ is reached for $t_{F}-t_{m}\simeq(q-1)(t_{m}-t_{c})$.
For $t>t_F$, the temperature is given by $T\simeq T_{c}-T_{c}H_c(t-t_{F})$.

\subsection{Semianalytic computations}

We have reduced the calculation of the phase transition dynamics to a relatively simple equation   for the average bubble radius $\bar R(t)$,  Eq.~(\ref{eqR}). Although this equation cannot be integrated analytically in the general case, it represents a considerable simplification for the numerical computation. 
We shall now compare its solution 
with the results of the more complete treatment of Sec.~\ref{dyn}. 

All the parameters appearing in  Eq.~(\ref{eqR}), 
as well as the initial condition for $\bar{R}$, 
can be estimated with the analytic and semi-analytic equations derived in Sec.~\ref{approxrate}.
We may use the initial condition $\bar{R}(t_m)=0$, 
which is consistent with the approximations (\ref{eqR}). We have checked that this is in very good agreement with the results of Sec.~\ref{dyn}. However,  an even better agreement
is obtained if we use as initial condition for the average radius the value given by Eq.~(\ref{rmexp}), $\bar{R}(t_{m})=v_{m}/\beta_m$. Similarly,  from Eqs.~(\ref{Iexp}) and (\ref{gamtrunc}) we may obtain an initial value for the fraction of volume,
\begin{equation}
I(t_{m})=\frac{3H_c}{r\beta_m}.\label{Im}
\end{equation}
These  values  of $\bar{R}$ and $I$ are consistent for an exponential nucleation rate but not for a delta-function
nucleation rate\footnote{That is, the quantities $I(t_m)$ and $\bar{R}(t_m)$
obtained with the sudden-reheating approximation will not fulfill, in general,
the relation $I=(4\pi/3)\bar{R}^{3}/d_f^{3}$ corresponding
to simultaneous nucleation.} and, hence, they will give a slightly different evolution. 
Here, we shall show the results for $f_+$ corresponding to the initial condition (\ref{Im}) and the results for $\bar R$ corresponding to the initial condition (\ref{rmexp}). 

\begin{figure}
\centering
\includegraphics[width=16cm]{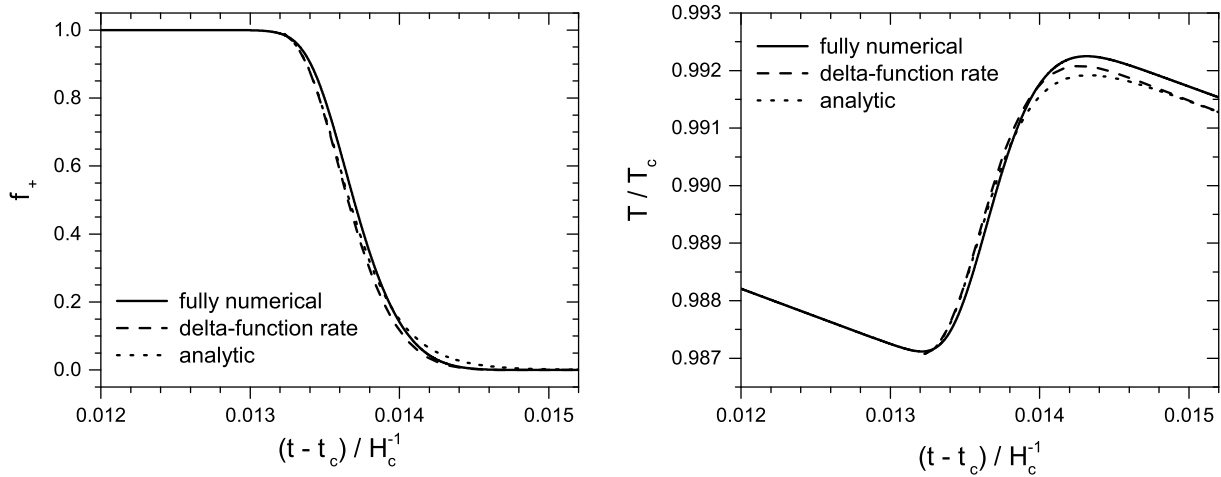}
\caption{Comparison of the numerical evolution of the phase transition computed in Sec.~\ref{dyn}
(solid lines) and different approximations, for the case $q<1$.
Dashed lines correspond to the numerical solution of the simplified equation (\ref{eqR}), and dotted lines correspond to the analytic approximation given
by Eq.~(\ref{tRcuart}).}
\label{figmath1}
\end{figure}

\begin{figure}
\centering
\includegraphics[width=16cm]{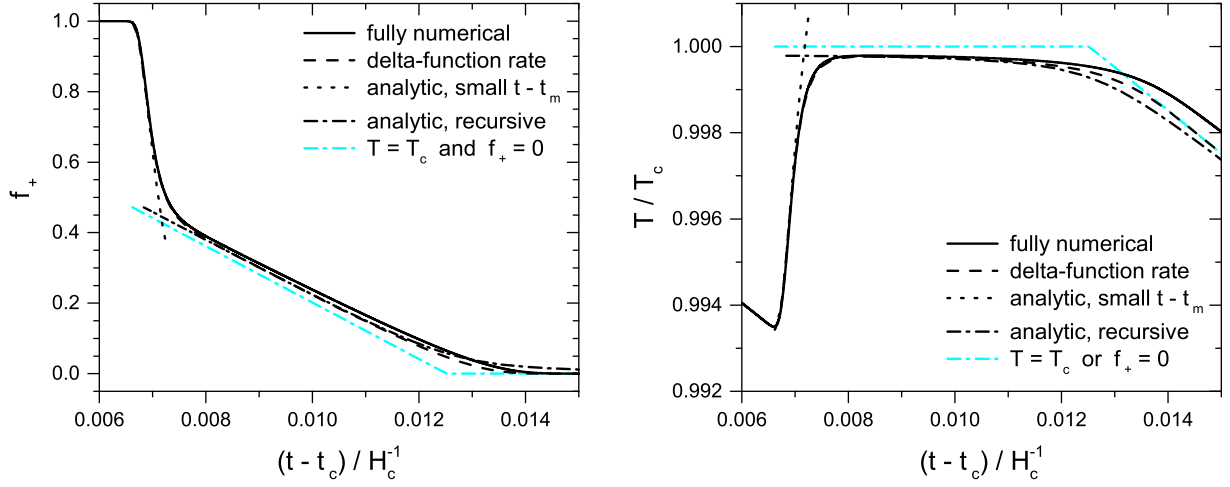}
\caption{Comparison of different computations of the evolution, for the case $q>1$. Solid,
dashed, and dotted lines are as in Fig.~\ref{figmath1}, cyan dashed-dotted lines correspond to the rough approximations $T=T_{c}$ and $f_{+}=0$, and black dashed-dotted lines correspond to
the improvement (\ref{tIrecur}).}
\label{figmath2}
\end{figure}

In Figs.~\ref{figmath1} and \ref{figmath2} we plot the fraction of volume and the temperature.  
The solid curves are those of Figs.~\ref{figtempvelo} and \ref{figevol}, while the dashed curves correspond to the solution of Eq.~(\ref{eqR}). The  other curves correspond to analytic approximations described below. 
We see that the simultaneous nucleation is
a very good approximation in these cases. We remark that we have used also the analytic approximations of  Sec.~\ref{approxrate} for the parameters in the equation and the initial conditions, which  introduce errors as well. As expected, the maximum error occurs
at the end of the phase transition. This is better appreciated in the temperature curves. 

\begin{figure}
\centering
\includegraphics[width=16cm]{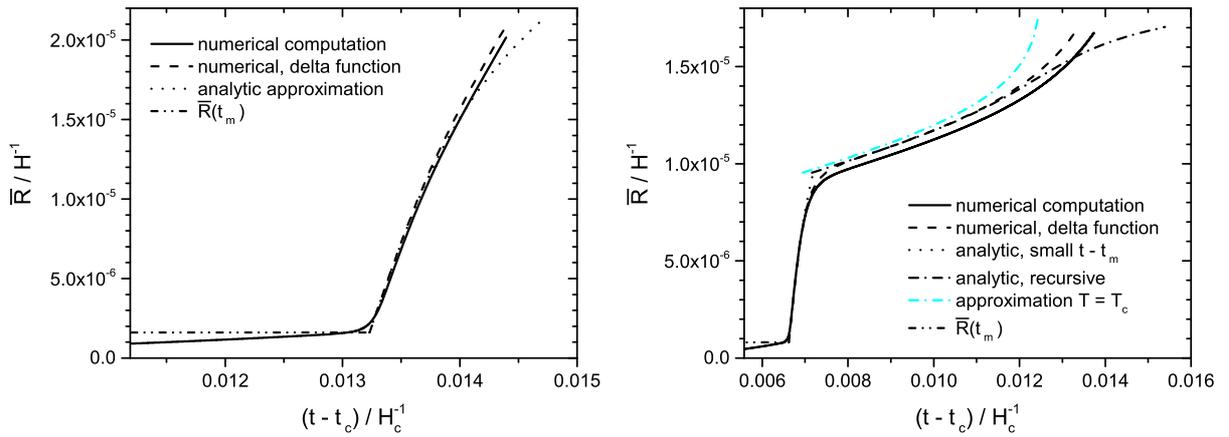}\caption{Average bubble radius as a function of time. The left panel corresponds to the case $q<1$ and the right panel to the case $q>1$.}
\label{figrmath}
\end{figure}

In Fig.~\ref{figrmath} we show the evolution of $\bar{R}$. 
The  numerical computation described in Sec.~\ref{dyn} (solid line) is  plotted  from $t=t_{N}$ to $t=t_{F}$. 
The dashed line is the solution of Eq.~(\ref{eqR}) and is plotted from $t=t_m$ to $t=t_F$. The time $t_F$ is different in each curve, since it is defined by $f_{+}(t_{F})=0.01$ for each calculation.

\subsection{Analytic approximations}

For $t<t_m$, we have $f_-=0$, and Eq (\ref{Tqtau}) is equivalent to $T=T_c-T_cH_c(t-t_c)$. 
For the average radius we have $\bar R\simeq 0$, and we may also use the value
$\bar{R}(t)\simeq\bar R(t_m)$, which is given by Eq.~(\ref{rmexp}) and is indicated in Fig.~\ref{figrmath} by a dash-dot-dot line.

To obtain analytic results for $t>t_m$, we must make further approximations to Eq.~(\ref{eqR}).
As already discussed, for the case $q<1$, 
neglecting the difference $t-t_{m}$  could
be a good approximation  until the end of the phase transition. 
In this case, we have $v_mdt/{d\bar{R}}=[1-q(1-e^{-I(\bar{R})})]^{-1}$ and,
for small $I$, we obtain $v_mdt/d\bar{R}=(1+qI)$, which can be readily integrated. We have
\begin{equation}
(t-t_{m})=\frac{1}{v_{m}}\left[\bar{R}-\bar{R}_{m}+\frac{\pi q}{3}\frac{\bar{R}^{4}-\bar{R}_{m}^{4}}{d_f^{3}}\right], \label{tRcuart}
\end{equation}
where $\bar{R}_{m}=\bar{R}(t_{m})$.
The function $\bar{R}(t)$ is thus obtained by inverting this quartic polynomial.
The approximation (\ref{tRcuart}) 
will break down for $I\gtrsim1$. Nevertheless, in this case we  have
$f_{+}\simeq0$, so the error will be irrelevant for several quantities, such as the temperature or the velocity. Indeed, in Eqs.~(\ref{Tqtau}) and (\ref{vwqtau}), at late times we have $f_-\simeq 1$ and the evolution is given by the last terms. The curves are shown with a dotted line in
Fig.~\ref{figmath1}. We
see that, indeed, this analytic approximation is very close to the numerical result.
The value of $\bar R$ for this case is shown in the left panel of Fig.~\ref{figrmath} (dotted line). As expected, it departs from the numerical result only at the end of the phase transition.
Notice that, using $\bar{R}=(\frac3{4\pi}I)^{1/3}d_f$, analytic approximations can be readily  obtained for the reference times $t_{I}$, $t_{P}$, $t_{E}$ and $t_{F}$
defined in Sec.~\ref{dyn}.

On the other hand, for $q>1$, we should not expect this approximation to remain valid
until the end of the phase transition, since the approximation of neglecting the difference $t-t_m$ in Eq.~(\ref{eqR}) breaks down.
Indeed, in Fig.~\ref{figmath2} this analytic approximation (dotted lines)
departs from the numerical computation as soon as the phase transition slows down. 

For the subsequent slow stage, we may use the rough approximation $T=T_{c}$,
which gives the linear function of Eq.~(\ref{fslow}).
This approximation, which is indicated with dash-dotted cyan lines in
Figs.~\ref{figmath2} and \ref{figrmath}, can be used until $f_+$ vanishes, where it can be matched to the approximation $f_{+}=0$.  
Although this rough approximation reproduces
quite well the behavior of $f_{+}$ and $T$, it is not useful for
the estimation of the wall velocity, since it corresponds to $v_{w}=0$. 
A better approximation can be obtained with a recursive trick. 
Notice that Eq.~(\ref{fslow}) is equivalent to $f_{-}=1/q+(3H_c/r)(t-t_{m})$,
which could have been obtained directly from Eq.~(\ref{tTsimple}) with the condition 
$\dot T\simeq 0$, which gives  $r\dot{f}_{-}=3H_c$ (the balance between the injection and extraction of energy from the plasma). In terms of $\bar R$, this condition is
$4\pi\dot{\bar{R}}\bar{R}^2 f_+/d_f^3=3H_c/r$. Inserting it on
the left-hand side of Eq.~(\ref{eqR}),  we obtain the  analytic relation
\begin{equation}
H_c(t-t_{c})=\frac{r}{3}(1-e^{-I})+\frac{1}{3}\left(\frac{3}{4\pi}\right)^{1/3}\frac{H_cd_f}{v_{m}}\frac{e^{I}}{qI^{2/3}}.\label{tIrecur}
\end{equation}
This relation gives the black dash-dotted curves in Fig.~\ref{figmath2} and in the right panel of Fig.~\ref{figrmath}. 
Notice that the approximations are very good, except near the end
of the phase transition. In particular, the value of $t_{F}$ for
the analytic approximation has a relatively large error.

\section{Implications for cosmic relics \label{cosmo}}

Although it  is out of the scope of this paper to
compute the possible relics from a phase transition, we wish to discuss on 
the implications of the dynamics we have just studied for their formation mechanisms.
Several simplifications  are often used in the literature. As already mentioned,
the most common approximations are a constant wall velocity and an exponential nucleation rate, and
the results for the remnants of the phase transition depend on the free parameters $v_w$ and $\beta_*$. 
In the case of small $v_w$, the dynamics also depends on a few parameters such as $v_m$ and $\beta_m$.
In this case, however, the computation of cosmic remnants will be more involved, due to
the non-trivial variation of the quantities.
Below we consider two of the possible relics and discuss on 
the computation
of the relevant quantities which are involved in their formation.

\subsection{Topological defects }

An important possible consequence of a cosmological phase transition
is the formation of topological defects (see \cite{defects} for reviews). We shall consider for concreteness
the case of cosmic strings. The simplest scenario in which these objects
may arise is that in which a global $U(1)$ symmetry is spontaneously
broken at the phase transition. Inside each bubble, the phase angle
$\theta$ of the Higgs field takes different values, and, when the
walls of two bubbles collide, $\theta$ interpolates smoothly between
these values \cite{k76}. As three or more bubbles meet, a total phase
equilibration may not be possible due to topological obstruction.
In such a case, the phase will change by $\Delta\theta=2\pi$ around
the line at which the bubbles meet. At this line the Higgs field vanishes,
and a cosmic string is formed. 

In the case
of a gauge symmetry, the phase difference is gauge dependent. One way of dealing
with this issue is to use a gauge invariant phase $\Delta\theta$, defined
as the line integral of the covariant derivative $D_{\mu}\theta=\partial_{\mu}\theta+eA_{\mu}$
\cite{kv95}. In this case, during phase equilibration a magnetic
flux is generated in the false vacuum region near the intersection
of two colliding bubble walls. When a third bubble arrives, the fluxes
corresponding to each pair of bubbles combine and, if there is a total
phase change of $2\pi$, a flux quantum is trapped inside the string.
An additional mechanism of string formation is due to the presence
of magnetic fields before the phase transition, which can be produced
by thermal fluctuations \cite{hr00}. After the phase transition,
this magnetic field will be trapped in quantized flux tubes. If the
phase transition is quick enough, this mechanism may produce a larger
density of strings \cite{hr01}, or strings with higher winding number
\cite{dr06}.

In either case, by the end of the phase transition, a random network
of cosmic strings  with some characteristic
length scale $\xi$ is expected to be formed. The statistical properties of such a network
were studied in numerical simulations with cells of size $\xi$ \cite{vv84,hs95,bov07}.
These calculations give, for instance, the proportions of closed loops
and infinite strings. 
However, the characteristic length $\xi$ is
in principle given by the separation between nucleation centers, which
is not a constant. 
In the first place, bubbles nucleate at random points, so the 
separation between neighboring bubbles has a dispersion $\Delta d$ around
its average $\bar d$, with $\Delta d\sim \bar d$. 
In the second place, bubbles nucleate
at different times, and those which were nucleated at the beginning of
the phase transition are larger than those which were nucleated near
the end. As a consequence, we will have inhomogeneities in the average separation $\bar d$.
In particular, for an exponentially growing nucleation rate, 
regions which were converted later to the broken-symmetry
phase contain a much larger number density of bubbles.
In contrast, for a slow phase transition, all bubbles nucleate in a relatively short time, and we expect a homogeneous average separation.

Similarly, for the bubble size distribution, in the exponential nucleation case we have
a dispersion $\Delta R\sim\bar{R}$, with $\bar R\simeq d_f\sim v_{w}\beta_*^{-1}$ at the end of the phase transition, 
while for the case of slow deflagrations the dispersion is quite smaller. Indeed, 
according to Eq.~(\ref{gdist}), we have $\Delta R\simeq v_{m}\beta_{m}^{-1}$,
while the final size $\bar{R}(t_{F})\simeq d_f$ is a factor $\sim\left(\beta_{m}/H\right)^{1/3}$
larger. 

Even for a single scale $d$, the final characteristic length depends on the dynamics of phase equilibration or magnetic field diffusion, and hence on that of bubble nucleation and growth. If 
the phase angle in each bubble is uniformly distributed between $0$ and $2\pi$,  the probability
of trapping a string between three bubbles is $1/4$. This gives a string length density
of order $1/(4d^{2})$. If bubbles expand at approximately the speed
of light, this is a good approximation, 
but for $v_{w}\ll1$,
phase equilibration between two collided bubbles may complete before
the wall of a third bubble reaches the meeting point, thus reducing
the probability of trapping a string. For a gauge theory, the evolution
of the phase difference is related to the spreading of magnetic
flux \cite{kv95}, and the conductivity plays a role in the process.
When two bubble walls collide, the magnetic flux generated at their
intersection will spread in the symmetric phase. If part of the flux
escapes to distances greater than the bubble radius before a third
bubble arrives, the probability of defect trapping will be suppressed.
Regarding the mechanism of flux trapping of already existing magnetic
fields, it has not been much investigated for first-order phase transitions.
Nevertheless, it is clear that the density of defects will be
smaller for slower phase transitions.

Different kinds of simulations have been performed (mainly in 2+1
dimensions; see e.g. \cite{bkvv95,f98,lf01}) to study the dependence
of defect formation on the dynamics of the phase transition. In these simulations, a
constant wall velocity as well as a constant nucleation rate were
assumed. As already discussed, the latter is generally a bad approximation.
If we assume that such a constant rate
turns on at a certain time $t_{0}$ and then takes
a  value $\Gamma(t)=\Gamma_{0}$, we obtain a
fraction of volume $f_{+}(t)=\exp[-(\pi/3)\Gamma_{0}v_{w}^{3}(t-t_{0})^{4}]$ (assuming also a constant $v_w$).
The final size distribution,  $dn/dR=(\Gamma_{0}/v_{w})f_{+}(t_{F}-R/v_{w})$,
is maximal at $R=v_{w}(t_{F}-t_{0})$. This result is qualitatively different from both the detonation and the slow deflagration cases. 

For the deflagration case, a simultaneous nucleation at $t=t_m\simeq t_I$ is a good approximation and is simpler than a constant rate.
Unfortunately, in this case
$v_w$ changes during the phase transition.
Nevertheless, the dynamics is simplified by the fact that all the bubbles have similar sizes, and our analytic approximations may be useful in the calculation.
Without entering into the details of the formation mechanisms, we notice that, although 
a common feature seems to be that smaller wall velocities reduce the probability of trapping defects in bubble collisions,  the bubble separation is also smaller for lower velocities, $d\sim v_m\beta_m^{-1}$, so the characteristic time between successive collisions, $\delta t\sim\beta_m^{-1}$, is rather independent of the wall velocity.

\subsection{Electroweak baryogenesis and baryon inhomogeneities}

The generation of the baryon asymmetry of the universe (BAU) may occur in the
electroweak phase transition  (see
\cite{mr12} for a review). The mechanism 
requires a first-order phase transition. In front of the walls of
expanding bubbles, chiral asymmetries in particle number densities
are generated due to C- and CP-violating scattering processes at the
interfaces. These asymmetries bias the baryon-number violating processes
(the sphalerons) in the symmetric phase. A net baryon number density
is thus formed and enters the bubbles, where baryon number violation
is turned off. A successful electroweak baryogenesis requires sufficient
CP violation as well as a strong enough phase transition. The latter
requirement is expressed quantitatively by the condition
 $\phi_{-}(T)/T\gtrsim1$, which guarantees
that sphaleron processes are suppressed in the broken-symmetry phase,
thus avoiding the washing out of the generated BAU. 

This mechanism
has also an important dependence on the wall velocity. The chiral
densities formed in front of the walls will be larger for higher velocities.
However, the walls must also be slow enough for the sphalerons to
have enough time to produce baryons. As a result, the generated
BAU peaks for a certain wall velocity $v_{w}=v_{\mathrm{peak}}$,
which depends on the interaction rates and diffusion constants, and
is generally in the range $10^{-2}\lesssim v_{\mathrm{peak}}\lesssim10^{-1}$
(see, e.g., \cite{ck00,cjk00,hs01,hjs01,cmqsw01}). 

In general, computations of electroweak baryogenesis for specific
models focus on the sources of CP violation and on the condition $\phi/T>1$,
and assume some (fixed) value for the wall velocity. On the
other hand, the velocity of the electroweak bubble wall has been 
investigated for several models (see, e.g., \cite{mp95,js01,knr14,k15,ms10}). Such computations generally focus on the determination of the friction of the wall with the plasma, taking into account
the hydrodynamics of an isolated bubble. The resulting $v_w$ depends on the temperature $T$ outside the bubble, which for application to specific models is usually evaluated at the onset
of nucleation, $T=T_{N}$. 

As we have seen, the wall velocity may vary significantly after the time $t_N$,
especially if its initial value is in the range
which is favorable for baryogenesis (as in our numerical examples).
Depending on the model, the effect of this velocity decrease may be either an enhancement \cite{h95} or a suppression \cite{m01} of the generated  baryon number density $n_B$. 
Indeed, for $v_w>v_{\mathrm{peak}}$ we have roughly  $n_{B}\propto v_{w}^{-1}$, while for $v_w<v_{\mathrm{peak}}$ we have roughly $n_{B}\propto v_{w}$.
Thus, if the initial velocity is  lower than $v_{\mathrm{peak}}$, the decrease of  $v_{w}$ will cause a suppression, while if the initial velocity  is (sufficiently) larger than $v_{\mathrm{peak}}$, we will have an enhancement.

In either case, a consequence of the velocity variation during baryogenesis
is the formation of baryon inhomogeneities \cite{h95,ma05}, due to a varying baryon
number density which is left behind by the moving walls.
A spherically-symmetric  density profile  is formed inside each bubble (at least, until bubbles meet each other).
Since all bubbles nucleate almost simultaneously (at $t=t_{m}$), the inhomogeneities 
will have similar sizes and profiles. A density $n_B(v_w(t))$ is generated at a distance $r=R(t_{m},t)$ from the bubble center. At the bubble center the value is approximately given by $n_B(v_m)$, and  
at a distance $R(t_m,t)$ there will be either an enhancement or a suppression by a factor $v_w(t)/v_m$ (if $v_m$ is far enough from $v_{\mathrm{peak}}$). Hence, since $R(t_m,t)\simeq\bar R(t)$,
the profile $n_B(r)$ inside the bubble is essentially given by the parametric curve of 
$v_{w}(t)/v_{m}$ vs.\ $\bar{R}(t)$.

\begin{figure}
\centering
\includegraphics[width=16cm]{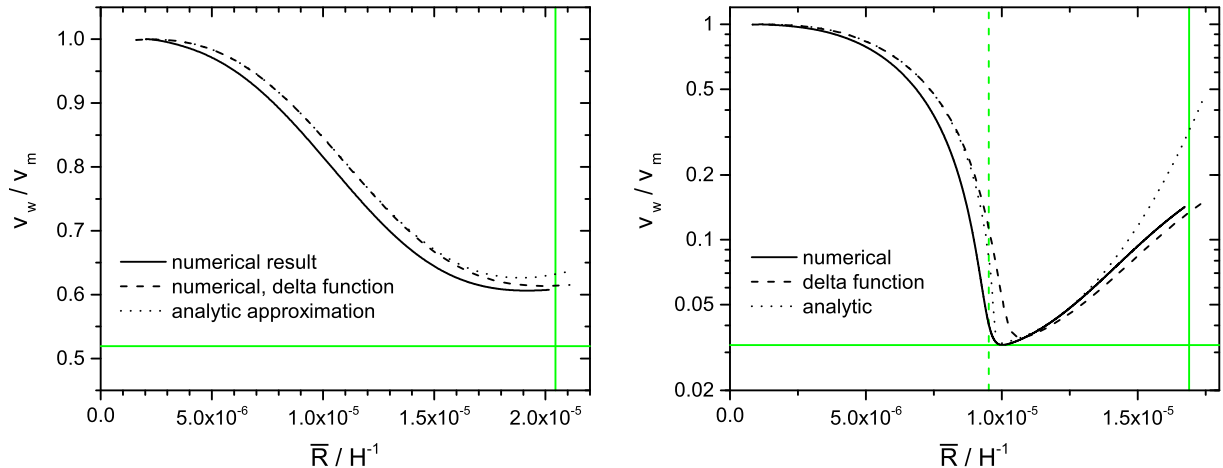}
\caption{The wall velocity vs.\ the average bubble radius between $t=t_{m}$
and $t=t_{F}$, for the cases $q<1$ (left panel) and $q>1$ (right
panel). Horizontal lines indicate  analytic estimations of
the minimum velocity. The solid vertical lines indicate the analytic estimation
of  $d_f$. The
dashed vertical line in the right panel corresponds to the estimation
of the moment at which $f_{-}=1/q$.}
\label{figvmath}
\end{figure}

The effect of reheating on the BAU and the formation of baryon inhomogeneities 
were investigated numerically in Refs.~\cite{h95,m01,ma05,ms08}. 
In Fig.~\ref{figvmath} we plot the wall  velocity  vs.\ the average bubble radius, which gives an idea of the inhomogeneity profile, for our numerical, semi-analytic, and analytic results. 
Solid lines correspond to the complete numerical computation,
 dashed lines correspond to the delta-function rate, and
 dotted lines correspond to the analytic approximations. In the latter case, the curves are easily obtained by inserting either relation
(\ref{tRcuart}) or (\ref{tIrecur}) in Eq.~(\ref{vwqtau}). 
The curves are plotted 
from $t=t_{m}$ to $t=t_{F}$. 
The approximation $R(t_{m},t)\simeq\bar{R}(t)$
is not valid for $t\simeq t_{m}$. At this time there is already a non-vanishing average
radius $\bar{R}(t_{m})\simeq v_{m}\beta_m^{-1}$, although most bubbles nucleate
at $t=t_{m}$. 
Like in Fig.~\ref{figrmath}, we observe that the analytic approximations
depart from the numerical results near $t=t_F$. We
remark that the  significant error observed in the right panel of Fig.~\ref{figvmath}
occurs in $\bar{R}(t)$ and not in $v_{w}(t)$, since 
Eq.~(\ref{tIrecur}) breaks down for small values of $f_{+}$,
and therefore does not have an impact on  Eq.~(\ref{vwqtau}).

It is useful to find also simple formulas  for the essential features, such as the amplitude of the inhomogeneities and their characteristic size.
In Ref.~\cite{ma05}, these basic characteristics were obtained  as functions of model parameters, using rough approximations such as
the analytic nucleation rate (\ref{nucanal}) and the estimation $T_{m}\simeq T_{N}$. 
We shall now obtain expressions in terms of the quantities $v_{m}$, $\beta_{m}$, and $q$. 

The maximum size scale of the inhomogeneities is the final bubble size $d_f$, which is given by Eq.~(\ref{dexp}).  This value is indicated by the solid vertical lines in Fig.~\ref{figvmath}.
However, the profile may have a shorter variation, depending essentially on  the ratio $q$ of Eq.~(\ref{qratio}). 
For $q<1$, the maximum temperature $T_r$ is reached at the end of the phase transition, so 
the minimum velocity $v_r$ is taken at the boundaries of the inhomogeneities. Therefore, the characteristic length of the profile is the whole length $d_f$. In this case,
we have a variation $v_r/v_{m}\simeq 1-q\sim 1$. This value of $v_r/v_m$ is indicated by the horizontal green line in the left panel of Fig.~\ref{figvmath}.

In the case $q>1$, 
the wall moves with velocity $v_{w}\sim v_{m}$ until
the fraction of volume becomes $f_{-}\sim1/q$. After that, we have
a much smaller velocity $v_r$ until the end of the phase transition. 
Thus, the baryon profile is  composed of two main parts;
namely, 
we have roughly a value $n_{B}(v_{m})$ inside a sphere of radius $\bar{R}_r$, while between this radius and $d_f$, we have roughly a value $n_{B}(v_{r})$. 
An estimate of the minimum velocity is obtained by setting $f_-=1/q$ in Eq.~(\ref{vwqtau}), which gives $v_{r}/v_{m}=(t_r-t_m)/(t_m-t_c)$. The corresponding time $t_r-t_m$ is given by Eq.~(\ref{tIrecur}) for $I=\log[q/(q-1)]$.
We thus obtain\footnote{The
parametric dependence $v_{r}\sim H_cd_f/r$ was already obtained in
Ref.~\cite{h95}.} 
\begin{equation}
v_{r}\simeq\left(\frac{3}{4\pi}\right)^{1/3}\frac{q/(q-1)}{\log[q/(q-1)]^{2/3}}
\frac{H_c d_f}{r},\label{vratio}
\end{equation}
This value is indicated by a green horizontal line in the right panel of Fig~\ref{figvmath}. 
The value  $\bar{R}_r$ corresponding to this value of $I$ is 
\begin{equation}
\bar{R}_r\simeq\left[\frac{3}{4\pi}\log\left(\frac{q}{q-1}\right)\right]^{1/3}d_f, \label{rinhom}
\end{equation}
indicated by a green dashed vertical line in Fig.~\ref{figvmath}.
In this approximation, the reheating temperature $T_r$ is given by $(T_r-T_m)/(T_c-T_m)=1-v_r/v_m$.
Taking into account that $r$ is proportional to $q$, if we vary this parameter we see that $v_r$ decreases roughly as $q^{-1}$.
On the other hand, the size $\bar R_r$ decreases more slowly.

\section{Conclusions \label{conclu}}

The dynamics of a phase transition which proceeds by the growth of
deflagration bubbles is quite different from that of a phase transition
mediated by detonation bubbles. The former is much more difficult
to study, due to the reheating caused by the release of latent heat. Nevertheless, the limit of very small velocities, $v_{w}<10^{-1}$,
is relatively simple, as the quantities 
depend on a homogeneous temperature $T(t)$. Thus, the distinctive characteristic
of the dynamics is a homogeneous reheating during the phase transition, which
causes the nucleation rate to turn off and the wall velocity to decrease. 
In spite of the aforementioned simplification, this scenario is still more complex than the detonation case. In the latter, an exponential nucleation rate and a constant wall velocity can be assumed, and the evolution of the phase transition can be solved analytically.
In contrast, in the slow-deflagration case, the nucleation rate, the wall velocity, and the temperature are linked through non-trivial equations.

This kind of phase transition has been considered in a number of works \cite{h95,m04,ma05,kk86,m01,ms08}. 
In Refs.~\cite{h95,ma05,m01,ms08} the development of the phase transition was computed numerically, while in Refs.~\cite{m04,kk86} some features, such as the minimum and maximum temperatures reached during the transition, were estimated analytically.
Only in the limit of a phase transition in equilibrium at $T=T_c$  the 
development can be solved exactly \cite{suhonen}. 
If the latent heat is large enough, the temperature eventually gets very close to $T_c$,
and the subsequent stage can be approximated by that limiting case.
However, this  phase-equilibrium stage is only a part of the evolution, and does not always occur.
The main aim of the present paper was to find analytic approximations for the nucleation rate, which allow to solve analytically the development of the phase transition in the general case, and to contrast the results with a complete numerical computation for a physical model.

During supercooling, the nucleation rate can be approximated by an exponential, like in the detonation case. However, when reheating begins bubble nucleation quickly turns off. With the approximation of a sudden reheating, we have found a simple semi-analytic equation for the minimum temperature $T_m$. Bubble nucleation occurs in a short interval around the corresponding time $t_{m}$. 
Assuming a constant wall velocity $v_w=v_w(T_m)$ in this interval, we obtained 
analytic approximations for several quantities,
such as the final average bubble separation (which is constant for $t\gtrsim t_m$). 
We have found that the quantities estimated at $t=t_m$ are in very good agreement with the numerical computation.
However, for the calculation of the bubble size distribution, a Gaussian nucleation rate (which gives a Gaussian distribution)  is more appropriate. 

In the evolution for $t>t_m$, the wall velocity can no longer be taken as a constant, and the simplest realistic approximation is a velocity of the form $v_w\propto T_c-T$, which 
is valid for a phase transition with little supercooling. 
Since the nucleation of bubbles is concentrated around $t=t_m$, a great simplification  is achieved by  considering  a nucleation rate of the form $\Gamma\propto\delta(t-t_{m})$. 
These approximations allowed us to obtain analytic solutions for the development of the phase transition after $t=t_m$. 
The effects of reheating are encoded in the parameter $q$ defined in Eq.~(\ref{q}).
For $q<1$, the temperature reaches a maximum $T_r\simeq T_m+q(T_c-T_m)$, and the velocity decreases only by a factor $1-q$. In this case, 
we have solved analytically the complete evolution. 
For the case $q>1$, this analytic solution only describes the reheating stage, after which the temperature gets very close to $T_c$ and the velocity can decrease by a few orders of magnitude. 
For this longer phase-equilibrium stage, we have found a refinement of the usual approximation in which the phase transition develops at $T=T_c$. 

These analytic solutions depend on a few parameters which can be evaluated at $t=t_m$.  We have verified that these approximations describe remarkably well the evolution of the 
quantities which are relevant for the generation of cosmic remnants.
This agreement with the numerical calculation shows in particular that
a delta-function nucleation rate is a good approximation. This is interesting since, in numerical simulations, the bubbles are sometimes nucleated simultaneously for simplicity. For this kind of phase transition, 
this approximation is more appropriate than using an exponential nucleation rate.

\section*{Acknowledgments}

This work was supported by FONCyT grant PICT 2013 No.~2786, CONICET grant
PIP 11220130100172, and Universidad Nacional de Mar del Plata, Argentina,
grant EXA793/16.

\appendix

\section{Entropy production and temperature variation}

We shall estimate the entropy increase during bubble growth.
Since we are assuming an ideal fluid, the entropy is conserved by the fluid equations, and it can only be produced in the discontinuities, i.e., in the bubble walls and in the shock fronts. For small wall velocities, the latter are extremely weak, and we can neglect these discontinuities. 
Therefore, we only consider the phase transition fronts. We may have overlapping bubbles, and we shall estimate the entropy produced at the  walls which remain uncollided at a given time (i.e., we consider the ``envelopes'' of bubble clusters). We assume that every surface element $\delta A$ moves with a velocity $v_w$ perpendicular to the surface. 

In the reference frame of a surface element, we assume that the fluid velocity is perpendicular to the wall. We assume deflagration conditions, in which the outgoing flow velocity $v_{-}$ is greater than the incoming velocity $v_+$.
Therefore, a portion of fluid which passes through the surface has a smaller entropy density but a larger volume 
in the $-$ phase.  In a time $\Delta t$ the entropy changes by 
\begin{equation}
\Delta S=(s_-v_--s_+v_+)\Delta t \delta A=\frac{v_-s_--v_+s_+}{v_-}v_w\Delta t\delta A,
\end{equation}
where we have assumed non-relativistic velocities, and we have used the deflagration relation $v_-=v_w$. Integrating over the uncollided wall area, we obtain 
\begin{equation}
\frac{dS}{dt}=\frac{v_-s_--v_+s_+}{v_-}\frac{df_-}{dt}V, \label{entprod}
\end{equation} 
where $V$  is the total volume\footnote{Thus, the fraction of volume in the $-$ phase is given by $V_-=f_-V$, and we have $dV_-=Av_wdt+f_-d V$, where $A$ is the total uncollided wall area. We also have $V\propto a^3$ and $\dot{V}/V=3H$.}. On the other hand, we have 
\begin{equation}
s\equiv S/V=s_-f_-+s_+f_+=s_+-(s_+-s_-)f_-, \label{smedia}
\end{equation} 
and Eq.~(\ref{entprod}) gives 
\begin{equation}
\dot s=\frac{v_-s_--v_+s_+}{v_-}\dot{f}_--3Hs. \label{entdensprod}
\end{equation}
If we neglect the entropy increase, Eq.~(\ref{entdensprod}) gives the right-hand side of Eq.~(\ref{varsmedia}), and Eq.~(\ref{smedia}) is the left-hand side. 

In order to compare the size of the different contributions to the temperature variation, it is convenient to differentiate (\ref{smedia}). We obtain
\begin{equation}
\frac{\dot{s}_+}{s_+}=\frac{s_+-s_-}{s_+}\dot{f}_--3H\left(1-\frac{s_+-s_-}{s_+} f_- \right)+\left(\frac{\dot{s}_+}{s_+}-\frac{\dot{s}_-}{s_+}\right)f_-+\left(\frac{s_-}{s_+}-\frac{v_+}{v_-}\right)\dot{f}_-.  \label{sdot}
\end{equation}
To convert this equation into an equation for the temperature, notice that  $\dot{s}_+/s_+=3\dot{T}_+/T_+$ and, from Eqs.~(\ref{discnr}), we have 
\begin{equation}
\frac{\dot{s}_-}{s_+}=\frac{s_-}{s_+}3\frac{\dot{T}_-}{T_-}=\frac{s_-}{s_+}\frac{v_-}{v_+}3 \frac{\dot{T}_+}{T_+}=\frac{T_+}{T_-}3\frac{\dot{T}_+}{T_+}.
\end{equation}
Using also the  relation $T_-s_-v_-=T_+s_+v_+$ in the last term of Eq.~(\ref{sdot}), we obtain
\begin{equation}
\left(1+\frac{T_+-T_-}{T_-}f_-\right)\frac{\dot{T}_+}{T_+}=
\frac{1}{3}\frac{s_+-s_-}{s_+}\dot{f}_-
-H\left(1-\frac{s_+-s_-}{s_+} f_-\right)
+\frac{1}{3}\frac{v_+}{v_-}\frac{T_+-T_-}{T_-}\dot{f}_-
\label{varT}
\end{equation}
We shall show that the two  terms proportional to $(T_+-T_-)/T_-$ can be generally neglected. 
We have also checked in our numerical computations that Eqs.~(\ref{rec}) and (\ref{varT})  do not give appreciable differences. The former corresponds to neglecting the last term in (\ref{varT})
(the entropy-production term).

In the first place, we have $(s_+-s_-)/s_+\sim L/\rho_{Rc}$. More precisely, from Eq.~(\ref{discnr}) we have
\begin{equation}
1-{s_-}/{s_+}=1-(1-3\alpha_c)^\frac{1}{4} (1-3\alpha_+)^\frac{3}{4}, \label{dsalfa}
\end{equation}
with  $\alpha_c\equiv\alpha(T_c)=L/(4\rho_{Rc})$ and $\alpha_+\equiv\alpha(T_+)=\alpha_cT_c^4/T_+^4$. Therefore, for small $L/\rho_{Rc}$ and $T_+\simeq T_c$, we may expand (\ref{dsalfa}) in powers of $\alpha$, and we obtain
\begin{equation}
\frac{s_+-s_-}{s_+}=\frac{3}{4}\frac{L}{\rho_{Rc}}
\end{equation}
plus terms of order $(L/\rho_{Rc})^2$ and $(L/\rho_{Rc})(T_c-T_+)/T_+$. In our numerical examples, these terms are of order $10^{-4}$, since we have  $L/\rho_{Rc}\sim(T_c-T_+)/T_+\sim10^{-2}$.
In the second term on the right-hand side of Eq.~(\ref{varT}) we may neglect the part of order $L/\rho_{Rc}$. In contrast, in the first term  we cannot do so, since $\dot{f}_-$ becomes much larger than $H$.  

On the other hand, we know that $(T_+-T_-)/T_-$  is small, and we may neglect it in the left-hand side of Eq.~(\ref{varT}). In contrast, in the right-hand side, the last term could be comparable to the first one. From Eq.~(\ref{discnr}) we have
\begin{equation}
\frac{T_+^4}{T_-^4}-1=\frac{3(\alpha_+-\alpha_c)}{1-3\alpha_+}=3\alpha_c\left(\frac{T_c^4}{T_+^4}-1\right)\left(1+\mathcal{O}(\alpha_+^2)\right). 
\end{equation}
Hence, to lowest order we obtain 
\begin{equation}
\frac{T_+-T_-}{T_-}=\frac{3}{4}\frac{L}{\rho_{Rc}}\frac{T_c-T_+}{T_+},
\end{equation}
and we may neglect the entropy-production term.  With these approximations, Eq.~(\ref{varT}) becomes
\begin{equation}
\frac{\dot{T}_+}{T_+}=
\frac{1}{4}\frac{L}{\rho_{Rc}}\dot{f}_- -H,
\end{equation}
which is equivalent to Eq.~(\ref{tTsimple}).


\begin{thebibliography}{99}

\bibitem{k76} T.~W.~B.~Kibble, 
J.\ Phys.\ A \textbf{9}, 1387 (1976). 

\bibitem{tw90} M.~S.~Turner and F.~Wilczek, 
Phys.\ Rev.\ Lett.\ \textbf{65}, 3080 (1990). 

\bibitem{krs85} V.~A.~Kuzmin, V.~A.~Rubakov and M.~E.~Shaposhnikov,   
Phys.\ Lett.\  {\bf 155B}, 36 (1985). 

\bibitem{w84} E.~Witten, 
Phys.\ Rev.\ D \textbf{30}, 272 (1984). 

\bibitem{h95} A.~F.~Heckler, 
Phys.\ Rev.\ D \textbf{51}, 405 (1995). 

\bibitem{tww92} M.~S.~Turner, E.~J.~Weinberg and L.~M.~Widrow,   
Phys.\ Rev.\ D {\bf 46}, 2384 (1992). 

\bibitem{nos}   A.~Megevand and S.~Ramirez,
  Nucl.\ Phys.\ B {\bf 919}, 74 (2017).
  
\bibitem{csw17}  R.~G.~Cai, M.~Sasaki and S.~J.~Wang,
  JCAP {\bf 1708}, no. 08, 004 (2017).
  

\bibitem{s82} P.~J.~Steinhardt, 
Phys.\ Rev.\ D \textbf{25}, 2074 (1982). 

\bibitem{gkkm84} M.~Gyulassy, K.~Kajantie, H.~Kurki-Suonio and L.~D.~McLerran, 
Nucl.\ Phys.\ B \textbf{237} (1984) 477. 

\bibitem{m04} A.~Megevand,   
Phys.\ Rev.\ D {\bf 69}, 103521 (2004). 

\bibitem{ma05} A.~Megevand and F.~Astorga,
Phys.\ Rev.\ D {\bf 71}, 023502 (2005). 

\bibitem{kk86} K.~Kajantie and H.~Kurki-Suonio, 
Phys.\ Rev.\ D {\bf 34}, 1719 (1986).  

\bibitem{m01} A.~Megevand, 
Phys.\ Rev.\ D {\bf 64}, 027303 (2001). 

\bibitem{ms08} A.~Megevand and A.~D.~Sanchez, 
Phys.\ Rev.\ D {\bf 77}, 063519 (2008). 

\bibitem{ah92} G.~W.~Anderson and L.~J.~Hall,
Phys.\ Rev.\ D \textbf{45}, 2685 (1992). 

\bibitem{affleck} I.~Affleck, 
Phys.\ Rev.\ Lett.\ \textbf{46}, 388 (1981). 

\bibitem{linde}
A.~D.~Linde, Phys.\ Lett.\ B \textbf{100}, 37 (1981); 
Nucl.\ Phys.\ B \textbf{216}, 421 (1983) [Erratum-ibid.\ B \textbf{223}, 544
(1983)]. 

\bibitem{lm16} L.~Leitao and A.~Megevand, 
JCAP {\bf 1605}, no. 05, 037 (2016). 

\bibitem {landau}L. D. Landau and E. M. Lifshitz, \textit{Fluid Mechanics}
(Pergamon Press, New York, 1989).

\bibitem{lm11}
  L.~Leitao and A.~Megevand,
  Nucl.\ Phys.\ B {\bf 844}, 450 (2011).

\bibitem{mp95} G.~D.~Moore and T.~Prokopec,
Phys.\ Rev.\ D \textbf{52}, 7182 (1995). 

\bibitem{ikkl94}  J.~Ignatius, K.~Kajantie, H.~Kurki-Suonio and M.~Laine,
  Phys.\ Rev.\ D {\bf 49}, 3854 (1994).

\bibitem{ekns10} J.~R.~Espinosa, T.~Konstandin, J.~M.~No and G.~Servant,
        JCAP {\bf 1006}, 028 (2010).

\bibitem{m13}  A.~M\'{e}gevand,
  JCAP {\bf 1307}, 045 (2013).

\bibitem{bm09}   D.~Bodeker and G.~D.~Moore,
  JCAP {\bf 0905}, 009 (2009).

  \bibitem{ms09}   A.~Megevand and A.~D.~Sanchez,
  Nucl.\ Phys.\ B {\bf 820}, 47 (2009).

\bibitem{k85} H.~Kurki-Suonio,
Nucl.\ Phys.\ B \textbf{255}, 231 (1985). 

\bibitem{gt80} A.~H.~Guth and S.~H.~H.~Tye,
Phys.\ Rev.\ Lett.\ \textbf{44}, 631 (1980) Erratum: [Phys.\ Rev.\ Lett.\
\textbf{44}, 963 (1980)]. 

\bibitem{gw81} A.~H.~Guth and E.~J.~Weinberg,
Phys.\ Rev.\ D \textbf{23}, 876 (1981). 

\bibitem{suhonen}   E.~Suhonen,
  Phys.\ Lett.\  {\bf 119B}, 81 (1982).

\bibitem{m06}  A.~Megevand,
  Phys.\ Lett.\ B {\bf 642}, 287 (2006).

\bibitem{jst17}  R.~Jinno, S.~Lee, H.~Seong and M.~Takimoto,
  arXiv:1708.01253 [hep-ph].

\bibitem{defects}  A. Vilenkin and E.P.S. Shellard, {\it Cosmic Strings and Other Topological Defects} (Cambridge University Press, Cambridge, England, 1994); A.~Vilenkin,  
Phys.\ Rept.\  {\bf 121}, 263 (1985);  
M.~B.~Hindmarsh and T.~W.~B.~Kibble, 
Rept.\ Prog.\ Phys.\  {\bf 58}, 477 (1995).   

\bibitem{kv95} T.~W.~B.~Kibble and A.~Vilenkin, 
Phys.\ Rev.\ D {\bf 52}, 679 (1995). 

\bibitem{hr00} M.~Hindmarsh and A.~Rajantie, 
Phys.\ Rev.\ Lett.\  {\bf 85}, 4660 (2000). 

\bibitem{hr01} M.~Hindmarsh and A.~Rajantie, 
Phys.\ Rev.\ D {\bf 64}, 065016 (2001). 

\bibitem{dr06} M.~Donaire and A.~Rajantie, 
Phys.\ Rev.\ D {\bf 73}, 063517 (2006) 

\bibitem{vv84} T.~Vachaspati and A.~Vilenkin,   
Phys.\ Rev.\ D {\bf 30}, 2036 (1984). 

\bibitem{hs95} M.~Hindmarsh and K.~Strobl, 
Nucl.\ Phys.\ B {\bf 437}, 471 (1995).

\bibitem{bov07} J.~J.~Blanco-Pillado, K.~D.~Olum and A.~Vilenkin, 
Phys.\ Rev.\ D {\bf 76} (2007) 103520. 

\bibitem{bkvv95} J.~Borrill, T.~W.~B.~Kibble, T.~Vachaspati and A.~Vilenkin, 
Phys.\ Rev.\ D \textbf{52}, 1934 (1995). 

\bibitem{f98} A.~Ferrera, 
Phys.\ Rev.\ D {\bf 57}, 7130 (1998). 

\bibitem{lf01} M.~Lilley and A.~Ferrera, 
Phys.\ Rev.\ D \textbf{64}, 023520 (2001). 

\bibitem{mr12}D.~E.~Morrissey and M.~J.~Ramsey-Musolf,   
New J.\ Phys.\  {\bf 14}, 125003 (2012). 

\bibitem{ck00} J.~M.~Cline and K.~Kainulainen, 
Phys.\ Rev.\ Lett.\  {\bf 85}, 5519 (2000). 

\bibitem{cjk00} J.~M.~Cline, M.~Joyce and K.~Kainulainen, 
JHEP {\bf 0007}, 018 (2000) 

\bibitem{hs01} S.~J.~Huber and M.~G.~Schmidt, 
Nucl.\ Phys.\ B {\bf 606}, 183 (2001). 

\bibitem{hjs01}S.~J.~Huber, P.~John and M.~G.~Schmidt, 
Eur.\ Phys.\ J.\ C {\bf 20}, 695 (2001). 

\bibitem{cmqsw01} M.~Carena, J.~M.~Moreno, M.~Quiros, M.~Seco and C.~E.~M.~Wagner,   
Nucl.\ Phys.\ B {\bf 599}, 158 (2001). 

\bibitem{js01} P.~John and M.~G.~Schmidt,
Nucl.\ Phys.\ B \textbf{598}, 291 (2001) [Erratum-ibid.\ B \textbf{648}, 449
(2003)]. 

\bibitem{knr14}
  T.~Konstandin, G.~Nardini and I.~Rues,
  JCAP {\bf 1409}, no. 09, 028 (2014).

\bibitem{k15}  J.~Kozaczuk,
  JHEP {\bf 1510}, 135 (2015).

\bibitem{ms10}  A.~Megevand and A.~D.~Sanchez,
  Nucl.\ Phys.\ B {\bf 825}, 151 (2010).



\end{thebibliography}
\end{document}